\documentclass[fleqn,usenatbib]{mnras}

\usepackage{newtxtext,newtxmath}

\usepackage[T1]{fontenc}
\usepackage{ae,aecompl}

\usepackage{graphicx}	
\usepackage{amsmath}	
\usepackage{amssymb}	
\usepackage{float}
\usepackage{algorithm}
\usepackage[noend]{algpseudocode}
\usepackage[FIGTOPCAP]{subfigure}

\def\f{\frac}

\def\ln{\mathrm{ln}}
\def\exp{\mathrm{exp}}
\def\cov{\mathbf{C}}

\def\data{\mathbf{d}}

\def\w{\mathbf{w}}

\def\btheta{\boldsymbol{\theta}}

\def\t{\mathbf{t}}

\title[Fast likelihood-free cosmology with neural density estimators]{Fast likelihood-free cosmology with neural density estimators and active learning}

\author[J. Alsing, T. Charnock, S. Feeney, B. Wandelt]{
Justin Alsing,$^{1,2,3}$\thanks{E-mail: justin.alsing@fysik.su.se}
Tom Charnock$^{4}$,
Stephen Feeney$^{2}$
and Benjamin Wandelt$^{2,5}$
\\
$^{1}$Oskar Klein Centre for Cosmoparticle Physics, Stockholm University, Stockholm SE-106 91, Sweden\\
$^{2}$Center for Computational Astrophysics, Flatiron Institute, 162 5th Ave, New York City, NY 10010, USA\\
$^{3}$Imperial Centre for Inference and Cosmology, Department of Physics, Imperial College London, Blackett Laboratory,\\  Prince Consort
Road, London SW7 2AZ, UK\\ 
$^{4}$Sorbonne Universit\'e, CNRS, UMR 7095, Institut d’Astrophysique de Paris, 98 bis bd Arago, 75014 Paris, France\\
$^{5}$Sorbonne Universit\'e, Institut Lagrange de Paris (ILP), 98bis boulevard Arago, F-75014 Paris, France\\
}

\date{Accepted XXX. Received YYY; in original form ZZZ}

\pubyear{2019}

\begin{document}
\label{firstpage}
\pagerange{\pageref{firstpage}--\pageref{lastpage}}
\maketitle

\begin{abstract}
Likelihood-free inference provides a framework for performing rigorous Bayesian inference using only forward simulations, properly accounting for all physical and observational effects that can be successfully included in the simulations. The key challenge for likelihood-free applications in cosmology, where simulation is typically expensive, is developing methods that can achieve high-fidelity posterior inference with as few simulations as possible. Density-estimation likelihood-free inference (DELFI) methods turn inference into a density estimation task on a set of simulated data-parameter pairs, and give orders of magnitude improvements over traditional Approximate Bayesian Computation approaches to likelihood-free inference. In this paper we use neural density estimators (NDEs) to learn the likelihood function from a set of simulated datasets, with active learning to adaptively acquire simulations in the most relevant regions of parameter space on-the-fly. We demonstrate the approach on a number of cosmological case studies, showing that for typical problems high-fidelity posterior inference can be achieved with just $\mathcal{O}(10^3)$ simulations or fewer. In addition to enabling efficient simulation-based inference, for simple problems where the form of the likelihood is known, DELFI offers a fast alternative to MCMC sampling, giving orders of magnitude speed-up in some cases. Finally, we introduce \textsc{pydelfi} -- a flexible public implementation of DELFI with NDEs and active learning -- available at \url{https://github.com/justinalsing/pydelfi}.
\end{abstract}
%



\section{Introduction}
\label{sec:introduction}
Likelihood-free inference (LFI) is emerging as a new paradigm for performing Bayesian inference under very complex generative models, using only forward simulations. This approach has great appeal for cosmological data analysis, since all effects that can be incorporated into forward simulations can be accounted for exactly in the inference pipeline, without having to resort to approximate calibrations and likelihood assumptions that may lead to biased inferences and/or mis-stated uncertainties. 

The main challenge for likelihood-free applications in cosmology, where simulation is expensive, has been developing methods that can give high-fidelity posterior inference from a feasibly small number of forward simulations. Traditional approaches to likelihood-free inference have been based on Approximate Bayesian Computation (ABC), which involves (variants on) drawing parameters from some proposal, simulating mock data, and accepting/rejecting the parameters based on whether the simulated data fall within some $\epsilon$-ball around the observed data (see \citealp{Lintusaari2017} for a review). Whilst ABC has enabled a number of applications in astronomy and cosmology \citep{Schafer2012,Cameron2012,Weyant2013,Robin2014,Lin2015,Hahn2017,Kacprzak2017,Carassou2017,Davies2017,Ishida2015, Akeret2015,Jennings2016}, ABC methods generally require a vast number of simulations, scaling exponentially with the number of model parameters, making them unfeasible when simulation is even modestly expensive.

Density-estimation likelihood-free inference (DELFI; \citealp{Bonassi2011,Fan2013, Papamakarios2016,Lueckmann2017,Papamakarios2018,Lueckmann2018,Alsing2018delfi}) aims to train a flexible density estimator for the target posterior from a set of simulated data-parameter pairs, and can yield high-fidelity posterior inference from orders-of-magnitude fewer simulations than traditional ABC-based methods. In this paper we introduce \textsc{pydelfi} -- a general purpose implementation of density-estimation likelihood-free inference using neural density estimators (NDEs) to learn the sampling distribution of the data as a function of the model parameters, employing active learning to adaptively run simulations in the most relevant regions of parameter space on-the-fly (based on \citealp{Papamakarios2018,Lueckmann2018}). We show that with NDEs and active learning, high-fidelity posteriors can be obtained for typical cosmological inference tasks from just a few thousand forward simulations. This opens up new possibilities for likelihood-free applications in cosmology.

The structure of this paper is as follows: In \S \ref{sec:delfi} we review density-estimation likelihood-free inference methods using neural density estimators and adaptive acquisition of simulations with active learning. In \S \ref{sec:compression} we review data compression schemes for accelerating likelihood-free inference; approximate score-compression, deep network parameter estimators, and information maximizing neural networks (IMNN; \citealp{Charnock2018}). In \S \ref{sec:implementation} we introduce \textsc{pydelfi}, briefly outlining the implementation details and features of the code. Tutorials and documentation for the code can be found at \url{https://github.com/justinalsing/pydelfi}. In \S \ref{sec:jla}--\ref{sec:Lya} we validate and demonstrate the performance of the \textsc{pydelfi} approach on some simple case studies from cosmology: analysis of the JLA supernova data \citep{Betoule2014} (against a known likelihood for validation), tomographic cosmic shear pseudo-$C_\ell$ analysis, and inference of the HI ionization rate around $z\sim 6$ from high-redshift Lyman-$\alpha$ forests. We conclude with some discussion in \S \ref{sec:conclusions}.
\section{Density estimation likelihood-free inference}
\label{sec:delfi}
In this section we provide a pedagogical review of density-estimation likelihood-free inference (\S \ref{sec:three_ways}) with neural density estimators (\S \ref{sec:ndes}-\ref{sec:ensemble}) and active learning to adaptively acquire simulations on-the-fly (\S \ref{sec:snl}). The methodology described in this section is based on \citet{Papamakarios2016}, \citet{Papamakarios2018}, \citet{Lueckmann2018} and \citet{Alsing2018delfi}.
\subsection{DELFI, three ways}
\label{sec:three_ways}
\begin{figure*}
\centering
\includegraphics[width = 17.5cm]{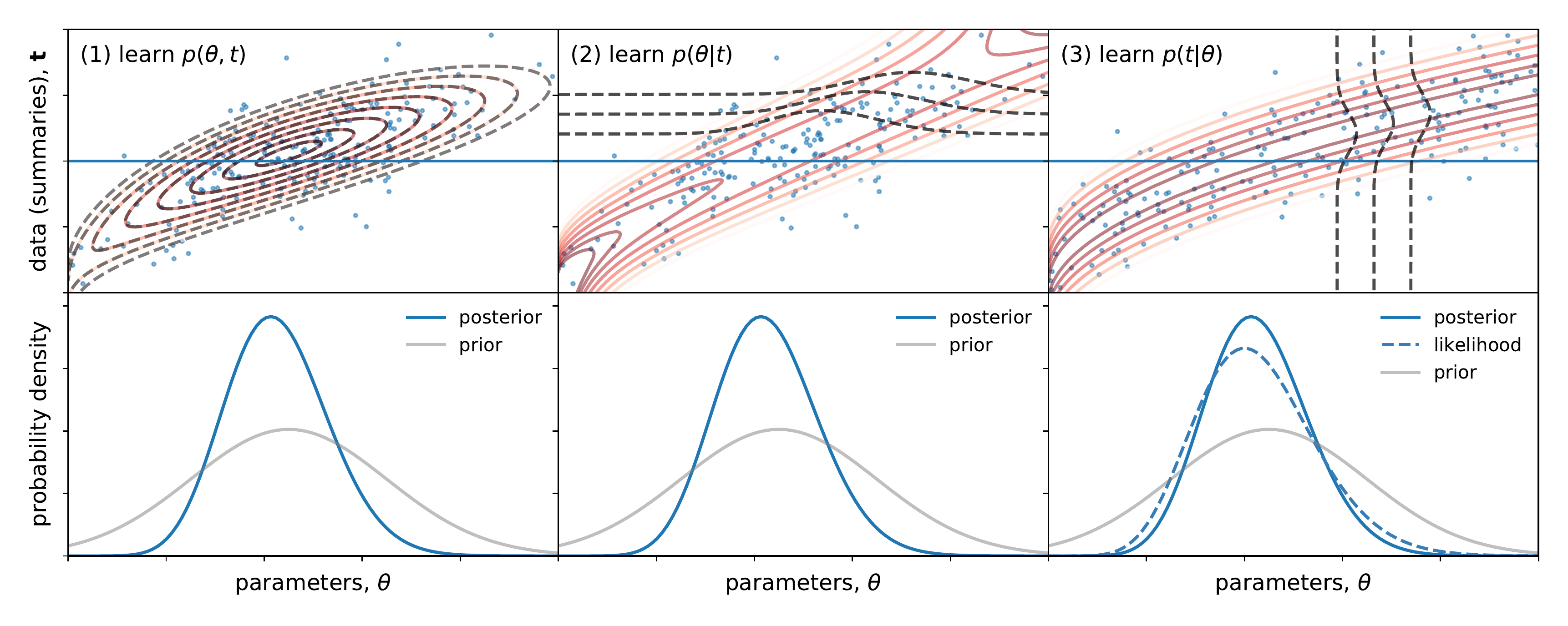}
\caption{Schematic for the three ways of performing density-estimation likelihood-free inference from a set of simulated data (summary) parameter pairs $\{\t,\btheta\}$: (1) learn a flexible parametric model for the joint density $p(\btheta,\t)$, (2) learn a flexible parametric model for the conditional density $p(\btheta | \t)$ (as a function of $\t$), (3) learn a flexible parametric model for the conditional $p(\t | \btheta)$ (as a function of $\btheta$). In each case, the goal is to learn the (conditional) density in the relevant region of parameter space, and take a slice at the observed data (summaries) to yield the target posterior or likelihood.}
\label{fig:3ways}
\end{figure*}
Density-estimation likelihood free inference turns inference into a density estimation task on a set of simulated parameter-data (summary\footnote{Throughout the text we use $\data$ to denote uncompressed data, $\t$ to denote compressed data summaries, and $\btheta$ to denote parameters. We write as though data are always compressed to some summaries $\t$ for likelihood-free inference, although this need not always be the case if the data are low-dimensional (relative to the number of simulations that can be performed -- see \S \ref{sec:compression} for discussion). We often use ``data" and ``data summaries" interchangeably in the text, being explicit where necessary to avoid confusion.}) pairs $\{\btheta, \t\}$. There are principally three ways to approach this density-estimation inference task (shown schematically in Figure \ref{fig:3ways}):

\begin{enumerate}
    \item[(1)] Fit a model to the joint density $p(\btheta, \t)$, then obtain the posterior by evaluating the joint density at the observed data $\t_o$, $p(\btheta | \t)\propto p(\btheta, \t=\t_o)$ \citep{Alsing2018delfi}.
    \item[(2)] Fit a model to the conditional density $p(\btheta| \t)$, then obtain the posterior by evaluating at the observed data $\t_o$. \citep{Papamakarios2016,Lueckmann2017}.
    \item[(3)] Fit a model to the conditional density $p(\t | \btheta)$, obtain the likelihood by evaluating at the observed data, and multiply by the prior to get the posterior $p(\btheta | \t)\propto p(\t | \btheta)\times p(\btheta)$ \citep{Papamakarios2018,Lueckmann2018}.
\end{enumerate}

Option 3 -- learning the sampling distribution of the data as a function of the parameters -- has some key advantages over the other two approaches. Firstly, by learning the sampling distribution of the data \emph{conditional} on the parameters, it does not matter how the parameters for running forward simulations were chosen. This gives complete freedom as to how simulations are acquired, so any schemes for adaptively acquiring simulations in the most relevant parts of parameter space can be employed without complication (see \S \ref{sec:snl}). In contrast, for options 1 and 2, parameters must either be drawn from the prior, or else drawn from some proposal density $q(\btheta)$ and the resulting learned target density subsequently re-weighted by $p(\btheta)/q(\btheta)$. This re-weighting step can result in instabilities during training, or high variance importance weights (and low effective sample sizes) after sampling, or both (see \citealp{Papamakarios2018} for discussion). By learning the likelihood function rather than the posterior, it is also more straightforward to explore different prior assumptions a posteriori without similar importance re-weighting issues.

Secondly, for applications where data are compressed to a small number of highly informative summaries, these will often tend to be asymptotically Gaussian, so for many problems the sampling distribution of the data summaries may be well-captured by a relatively simple density model (eg., a Gaussian mixture with a modest number of mixture components, or similar), even when the posterior (option 2) or joint distribution (option 1) is complicated.

In light of these considerations, we suggest implementing DELFI by learning the sampling distribution of the data (summaries) as a function of the model parameters as a sensible default approach\footnote{However, we note that option 1 comes with its own unique advantage in that it provides an analytical estimate of the Bayesian evidence for free, provided an analytically integratable joint-density parameterization such as a Gaussian mixture is used \citep{Alsing2018delfi}. This may be preferred when the evidence is the primary target. Note the evidence estimated this way will be with respect to the compressed summaries, rather than the un-compressed data vector.}. With this choice made, DELFI can be broadly summarized as follows:
\begin{enumerate}
\item Run simulations at different parameter values $\btheta$ to obtain simulated parameter-data pairs $\{\btheta, \t\}$, 
\item Fit a parametric conditional density estimator $p(\t | \btheta ; \w)$ to the simulations $\{\btheta, \t\}$, 
\item Evaluate the estimated conditional density at the observed data $\t_o$ to obtain the (learned) likelihood function $p(\t_o | \btheta ; \w)$.
\end{enumerate}
An efficient algorithm for performing DELFI must then address three key questions:
\begin{enumerate}
    \item How do we parameterize the conditional density estimator $p(\t | \btheta ; \w)$ in a sensible way?
    \item How do we run simulations in the most relevant parts of parameter space for the ultimate target, $p(\t_o | \btheta ; \w)$, to best use the available resources?
    \item If the uncompressed data vector $\data$ is high-dimensional, how can we compress it effectively to some small set of informative summaries $\data \rightarrow \t$ to reduce the dimensionality of the density-estimation task, and hence reduce the number of simulations required?
\end{enumerate}
In this paper we use neural density estimators (NDEs) as a flexible and efficient conditional density estimation framework for DELFI (based on \citealp{Papamakarios2016, Papamakarios2018, Lueckmann2018}), employing ensembles of networks (with different initializations and architectures) to give robustness against small training sets and architecture choice. We give an overview of NDEs and network ensembles in \S \ref{sec:ndes} and \ref{sec:ensemble}.

For efficient acquisition of simulations, we use active learning, allowing the NDEs to call the simulator to run new simulations on-the-fly, based on the current likelihood-surface approximation. We discuss active learning strategies in \S \ref{sec:snl}.

We review key data compression schemes for accelerating DELFI in \S \ref{sec:compression} (approximate-score compression, and deep network compression schemes).
\subsection{Neural density estimators}
\label{sec:ndes}
Neural density estimators (NDEs) provide flexible parametric models for conditional probability densities $p(\t | \btheta ; \mathbf{w})$, parameterized by neural networks with weights $\mathbf{w}$, which can be trained on a set of simulated data-parameter pairs $\{\t, \btheta\}$. 

In this section we review two classes of NDEs that have proven useful in the context of likelihood-free inference: mixture density networks (MDNs; \citealp{Bishop1994}) and masked autoregressive flows (MAFs; \citealp{Papamakarios2017}). Note this section assumes basic background knowledge of neural networks -- see eg., \citet{Bishop2006} for a comprehensive review.
\subsubsection{Mixture Density Networks (MDN)}
Mixture density networks constitute a class of models for the conditional density $p(\t | \btheta ; \mathbf{w})$ where the distribution for $\t$ at any given $\btheta$ is given by a mixture model, and the relative weights and properties of the mixture components are all free functions of $\btheta$, parameterized by a neural network with weights $\mathbf{w}$. For example, a Gaussian mixture density network\footnote{We will henceforth take MDN to mean Gaussian MDN (although other mixture models may be useful in certain situations).} defines the following conditional density estimator,
\begin{align}
p(\t | \btheta;\w) = \sum_{k=1}^{n_c} r_k(\btheta ; \w)\, \mathcal{N}\left[\t\,|\, \boldsymbol{\mu}_k(\btheta ; \w), \mathbf{C}_k\equiv\boldsymbol{\Sigma}_k(\btheta ; \w)\boldsymbol{\Sigma}_k^T(\btheta ; \w)\right],
\end{align}
ie., an $n_c$ component Gaussian mixture model where the component weights $\{r_k(\btheta ; \w)\}$, means $\{\boldsymbol{\mu}_k(\btheta ; \w)\}$, and covariance factors\footnote{To avoid redundancy from the positive-definiteness of the covariance matrices, it is practical if the neural network parameterizes only the (upper triangular) Cholesky factors of the component covariances.} $\{\boldsymbol{\Sigma}_k(\btheta ; \w)\}$ are all functions of $\btheta$ parameterized by a neural network with weights $\mathbf{w}$. 

\begin{figure*}
\centering
\includegraphics[width = 16.5cm]{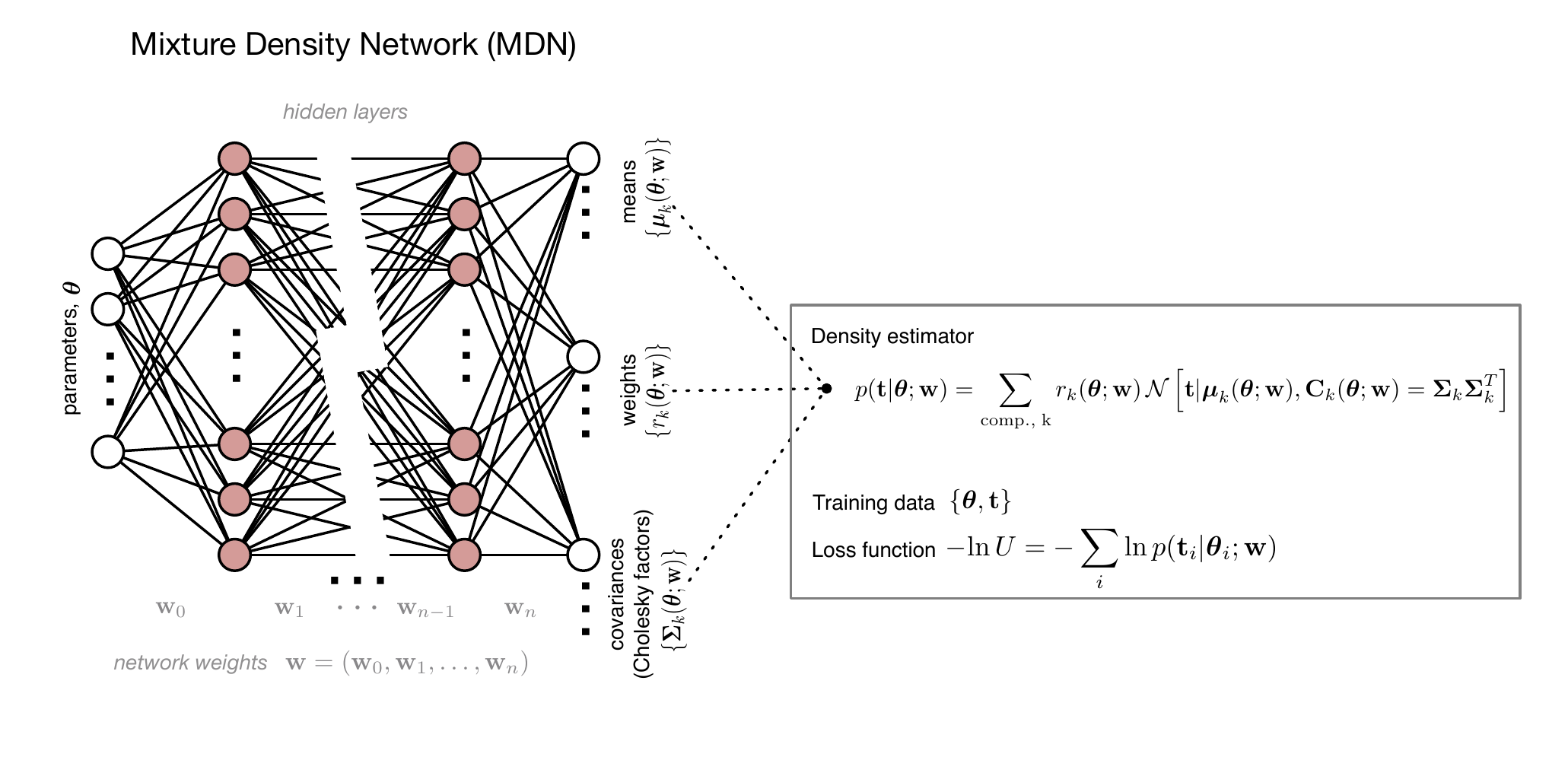}
\caption{Schematic of the mixture density network parameterization of the conditional density $p(\t | \btheta)$. The means, weights and covariances of a Gaussian mixture model for $p(\t | \btheta)$ are free functions of the parameters $\btheta$, parameterized by the weights, ${\mathbf{w}}$, of the neural network. The neural network takes $\btheta$ as input and outputs the parameters of the mixture model for those parameters.}
\label{fig:mdn}
\end{figure*}

The MDN model is shown schematically in Figure \ref{fig:mdn}; the network takes in parameters $\btheta$ and outputs the means, weights and covariances of the mixture model for $p(\t | \btheta)$ corresponding to that input $\btheta$. The MDN network architecture typically has a number of intermediate dense hidden layers with some non-linear activation function (eg., $\mathrm{tanh}$). In the output layer, the output nodes corresponding to the means have linear activations, as do the off-diagonal elements of the covariance matrices, whilst the diagonal covariance elements are passed through an exponential activation to ensure positive definiteness, and the mixture component weights are passed through a softmax activation\footnote{Softmax: $\mathbf{x}\rightarrow \exp(\mathbf{x})/\Sigma\,\exp(x_i)$. } to ensure they are positive and sum to unity.

Note that a mixture density network parameterization of $p(\t | \btheta)$ with a single Gaussian component defines a Gaussian likelihood where the mean and covariance are functions of the parameters -- a common approximate likelihood used in many cosmological data analysis problems. Adding additional components immediately results in a more flexible density estimator and hence likelihood assumptions; Gaussian mixtures can represent any smooth probability density (given enough components).
\subsubsection{Masked Autoregressive flows (MAF)}
\begin{figure*}
\centering
\includegraphics[width = 16.5cm]{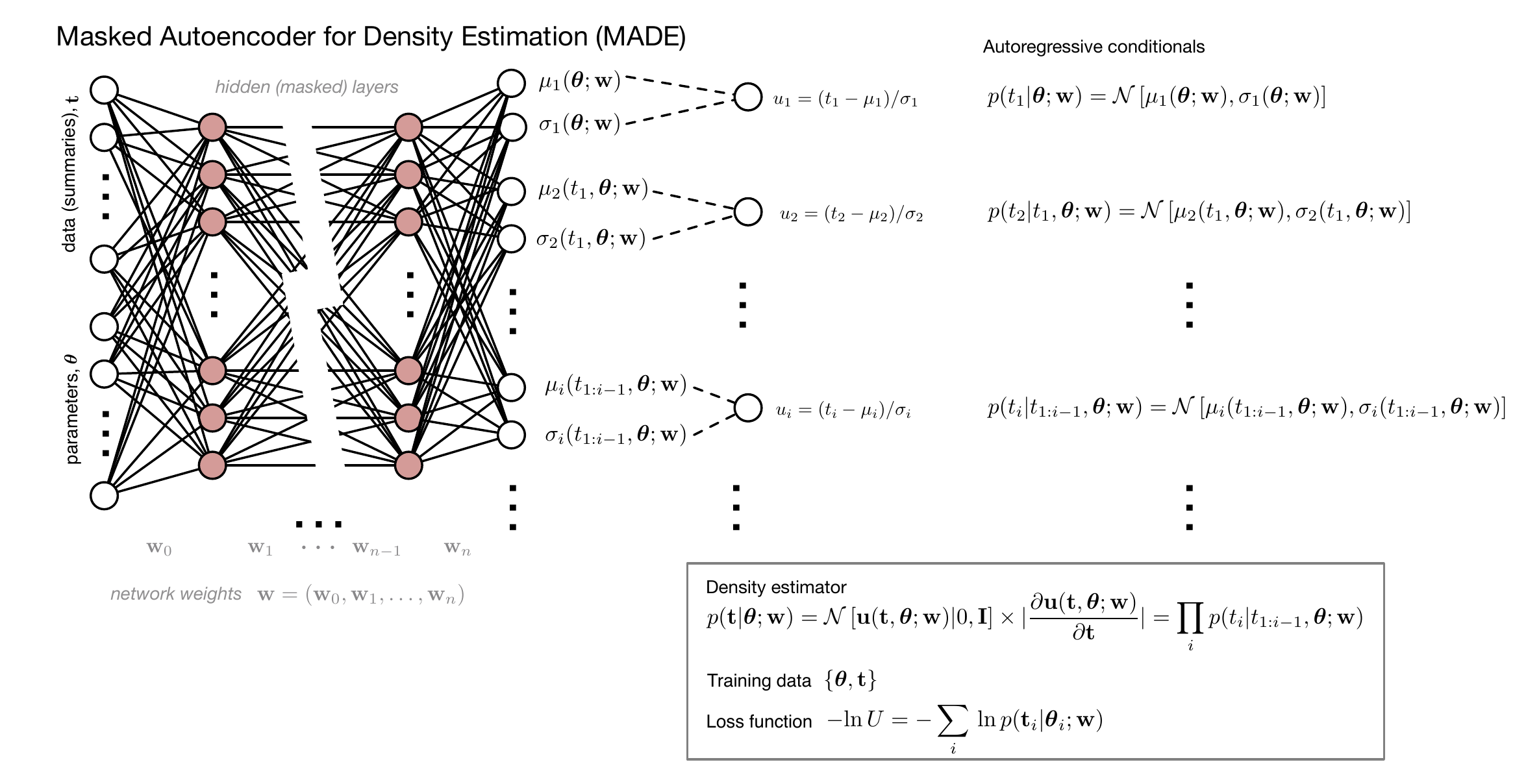}
\caption{Schematic of the conditional masked autoencoder for density estimation (MADE) parameterization of the conditional density $p(\t | \btheta)$. The means and variances of the autoregressive conditionals are parameterized by the neural network, with the hidden layers carefully masked to ensure the autoregressive properties are satisfied. A masked autoregressive flow (MAF) is a stack of MADEs, where the output of each MADE is fed as input to the next, and the order of the autoregressive factorization is changed between MADEs.}
\label{fig:made}
\end{figure*}

Any probability density can be factorized as a product of one-dimension conditionals via applications of the chain rule:
\begin{align}
\label{chain_rule}
    p(\t | \btheta) = \prod_{i=1}^{\mathrm{dim}(\t)} p(t_i | \t_{1:i-1}, \btheta).
\end{align}
Neural autoregressive density estimators construct parametric densities for this set of one-dimensional conditionals, where the parameters of each of the conditionals are parameterized as a neural network \citep{Uria2016}. For example, one could model each conditional $p(t_i | \t_{1:i-1}, \btheta)$ as a Gaussian whose mean and variance are free functions of $(\t_{1:i-1}, \btheta)$, parameterized by a neural network. Masked Autoencoders for Density Estimation (MADEs; \citealp{Germain2015}), depicted in Figure \ref{fig:made}, do precisely this: the means and variances of each conditional density are parameterized by the neural network, where crucially the weights of the neural network layers are masked in such a way that the output nodes for $p(t_i | \t_{1:i-1}, \btheta)$ only depend on $(\t_{1:i-1}, \btheta)$ (ie., the autoregressive property is preserved). See \citet{Germain2015} for details of how to construct the binary network weight mask. As with MDNs, the hidden layers of the MADE have some non-linear activation functions (eg., $\mathrm{tanh}$), whilst the output nodes associated with the conditional means have linear activation, and the output nodes associated with the variances have exponential activations (ensuring positivity).

By learning the means and variances of the autoregressive conditionals, a MADE can be thought of as learning the transform of the random variate $\t$ back to the unit normal:
\begin{align}
&\t | \btheta \rightarrow\mathbf{u}(\t, \btheta; \mathbf{w})\sim\mathcal{N}(\mathbf{0}, \mathbf{I}), \nonumber \\
&t_i | \btheta  \rightarrow u_i = (t_i - \mu_i(\t_{1:i-1}, \btheta; \mathbf{w}))/\sigma_i(\t_{1:i-1}, \btheta; \mathbf{w}),
\end{align}
where $\mathbf{w}$ are the (masked) weights of the neural network. The parametric density estimator for a MADE is hence given by,
\begin{align}
    p(\mathbf{t} | \boldsymbol\theta ; \mathbf{w}) &= \prod_i p(t_i | \t_{1:i-1}, \boldsymbol\theta;\mathbf{w}) \nonumber \\
    &= \mathcal{N}\left[\mathbf{u}(\mathbf{t}, \boldsymbol\theta;\mathbf{w}) | \mathbf{0}, \mathbf{I}\right] \times\left|\frac{\partial\mathbf{u}(\mathbf{t}, \boldsymbol\theta;\mathbf{w})}{\partial\mathbf{t}}\right| \nonumber \\
    &= \mathcal{N}\left[\mathbf{u}(\mathbf{t}, \boldsymbol\theta;\mathbf{w}) | \mathbf{0}, \mathbf{I}\right] \times\prod_{i=1}^{\mathrm{dim}(\t)}\sigma_i(\mathbf{t}, \boldsymbol\theta;\mathbf{w})
\end{align}
Single MADE density estimators have two key limitations. Firstly, they are sensitive to the order of the factorization in Eq. \eqref{chain_rule}; some densities may have simple (eg., unimodal) conditionals in one factorization-order, but not in another, and this is typically not known a priori (see \citealp{Papamakarios2017} for an illustration). Secondly, the assumption of simple (eg., Gaussian) conditionals may be overly restrictive.

Masked Autoregressive Flows (MAF; \citealp{Papamakarios2017}) address both of these limitations by constructing a stack of MADEs, where the output $\mathbf{u}$ of each MADE is taken as input for the next, with random re-ordering of the chain-rule factorization between each MADE. With multiple stacked MADEs and re-ordering, MAFs constitute very flexible neural autoregressive density estimators suitable for likelihood-free inference \citep{Papamakarios2018}. MAFs then define the following conditional density estimator:
\begin{align}
    p(\mathbf{t} | \boldsymbol\theta ; \mathbf{w}) &= \prod_i p(t_i | t_{1:i-1}, \boldsymbol\theta;\mathbf{w}) \nonumber \\
    &= \mathcal{N}\left[\mathbf{u}(\mathbf{t}, \boldsymbol\theta;\mathbf{w}) | \mathbf{0}, \mathbf{I}\right] \times\prod_{n=1}^{N_\mathrm{mades}}\prod_{i=1}^{\mathrm{dim}(\t)}\sigma_i^n(\mathbf{t}, \boldsymbol\theta;\mathbf{w}),
\end{align}
where $\mathbf{u}$ is the output from the final MADE.
\subsubsection{Training neural density estimators}
To fit a neural density estimator to a set of simulated samples $\{\btheta, \t\}$, we want to find the weights of the neural network that minimize the Kullback-Leibler divergence between the parametric density estimator $p(\t | \btheta ; \w)$ and the target $p^*(\t | \btheta)$:
\begin{align}
D_\mathrm{KL}(p^*\,|\, p ) = \int p^*(\t | \btheta)\;\mathrm{ln}\left(\frac{p(\t | \btheta;\w)}{p^*(\t | \btheta)}\right) d\t
\end{align}
Since we do not have access to the target density, only samples from it $\{\t, \btheta\}$, we take the (negative log) loss function to be:
\begin{align}
\label{nde_loss}
    -\ln\,U(\w | \{\btheta, \t\}) = -\sum_{i=1}^{N_\mathrm{samples}}\,\ln\,p(\t_i | \btheta_i;\w),
\end{align}
ie., a Monte Carlo estimate of the KL-divergence (up to an additive $\w$-independent constant), which is equivalent to the negative log-likelihood of the simulated data $\{\t, \btheta\}$ under the conditional density estimator $p(\t | \btheta ; \w)$.

For (Gaussian) MDN conditional density estimators, the loss is hence given by:
\begin{align}
    -\ln\,U(\w | \{\btheta, \t\}) = -\sum_{i}\,\sum_{k=1}^{n_c} \pi_k(\btheta_i ; \w)\, \mathcal{N}\left[\t_i\,|\, \boldsymbol{\mu}_k(\btheta_i ; \w), \boldsymbol{\Sigma}_k(\btheta_i ; \w)\right].
\end{align}

For MAF conditional density estimators, the loss is given by:
\begin{align}
    -\ln\,U(\w | \{\btheta, \t\}) = -\sum_{i}\,\ln\big[\mathcal{N}\big(\mathbf{u}(\t_i, &\btheta_i ; \mathbf{w})\, | \, \mathbf{0}, \mathbf{I}\big) \nonumber \\
    & + \sum_{n=1}^{N_\mathrm{mades}}\sum_{m=1}^{\mathrm{dim}(\t)}\ln\,\sigma_m^n(\t_i, \btheta_i ; \mathbf{w})\big].
\end{align}
The neural density estimators are then trained in the usual way by minimizing the negative log-loss with respect to the network weights, or inferring a posterior density over the weights given the training data (and some network weight prior). 

Over-fitting can be mitigated by any of the standard regularization methods used for neural networks, such as early-stopping or dropout. Early-stopping splits the training data into a training and validation set, and terminates training when the loss ceases to improve for the validation set. Dropout masks some subset of the hidden units, chosen at random, at each training iteration \citep{Srivastava2014}. Further regularization can be achieved using ensembles of networks, or Bayesian networks, as described in the next section.
\subsection{Bayesian networks, deep ensembles and stacked density estimators}
\label{sec:ensemble}
Whilst training sufficiently complex NDEs on sufficiently large training sets provides a robust approach to likelihood-free inference in practice \citep{Papamakarios2016, Papamakarios2018}, some simple sophistications can improve robustness to the choice of network architecture and small training sets (ie., when simulation is expensive).

Training individual NDEs (by minimizing the log-loss) on small training sets runs the risk of finding single local minima that may not best represent the training data. Furthermore, it may be difficult to choose an appropriate NDE network architecture for the problem at hand a priori; there is a trade-off to be made between ensuring sufficient model complexity to fit the unknown data distribution, whilst avoiding over-fitting.

One simple resolution is to train an ensemble of NDEs \citep{Lakshminarayanan2017, Lueckmann2018} $\{p_{\alpha}(\t | \btheta ; \mathbf{w})\}$, with a range of network architectures (and initializations). This ensemble can then be used to form a stacked density estimator for the learned likelihood-surface by stacking the individual trained NDEs,
\begin{align}
    p(\t | \btheta ; \mathbf{w}) = \sum_{\alpha=1}^{N_\mathrm{NDEs}} \beta_\alpha\,p_{\alpha}(\t | \btheta ; \mathbf{w}),
\end{align}
where the weights $\beta_\alpha$ are given by the relative likelihoods for each NDE, or their cross-validation scores \citep{Smyth1998,Smyth1999}. Stacked density estimators constructed this way are found to outperform a single best density estimator chosen from an ensemble of fits \citep{Smyth1998,Smyth1999}, and stacking ensembles of trained neural networks is common practice in machine learning for improving predictive accuracy.

As well as increasing robustness in the small training-set regime and against architecture choice, training ensembles of NDEs also allows for straightforward estimation of the uncertainty in the learned likelihood surface (ie., the weighted variance of the NDEs in the ensemble). This can be exploited in active learning schemes that use the uncertainty in the current likelihood-surface approximation to decide where to run new simulations, as described below.

A second approach to making neural networks robust in the small training-set regime is to train the networks in a Bayesian context, inferring a posterior distribution for the network weights given the training data, $p(\mathbf{w} | \{\t,\btheta\})$ (see eg., \citealp{Burden2008}). This helps to regularize the networks in two ways; firstly, it allows us to put a prior over the network weights, eg., imposing some sparse regularization. Secondly, the inferred network-weight posterior can be used to define an expectation value of the network output, and also to assess uncertainty in the network output that can be exploited in active learning schemes, in a similar spirit to ensembles of networks. Bayesian networks are reported to be robust to over-fitting and remove the need for additional over-fitting mitigation strategies.

Bayesian networks have the appeal over network ensembles that they allow the user to control the regularization in an interpretable way through the prior, and provide principled uncertainties and expectation values for the network output. Ensembles on the other hand have the advantage that they are trivial to implement; optimizing an ensemble of networks is typically simpler and cheaper in practice than inferring the posterior over the weights for a single large network (although fast approximate inference schemes such as variational inference help). Ensembles also make easy work of model averaging over different network architectures, which is more difficult (although still possible) in the Bayesian framework. Recent developments in Bayesian inference and subsequent marginalization over network architectures may prove useful in this context \citep{Higson2018, Dikov2019}.
\subsection{Adaptive acquisition of simulations with active learning}
\label{sec:snl}
When performing likelihood-free inference in situations where forward simulation is expensive, the goal is to achieve the highest fidelity posterior inference with the fewest simulations possible. We therefore want to preferentially run simulations in the most interesting regions of the parameter space, which are not known a priori. Active learning allows the neural density estimators to call the simulator independently during training, automatically deciding on-the-fly where the best parameters to run new simulations are based on their current state of knowledge/ignorance of the target posterior. Here we present two key active learning approaches for adaptive acquisition of simulations for DELFI: sequential neural likelihood (based on \citealp{Papamakarios2018}), and Bayesian optimization style acquisition rules (based on \citealp{Lueckmann2018}).
\subsubsection{Active learning with Sequential Neural Likelihood (SNL)}
The Sequential Neural Likelihood (SNL; \citealp{Papamakarios2018}) approach runs simulations in a series of batches, where the parameters for each batch of new simulations are drawn from a proposal density based on the current posterior approximation, and the NDEs are re-trained after each new simulation batch. This way, the algorithm adaptively learns the most relevant parts of the parameter space to run new simulations and hence improve the ongoing posterior inference.

How to define an optimal proposal density based on the current posterior approximation is an open question. \citet{Papamakarios2018} used the current posterior approximation directly as their proposal for new simulations, which is a natural choice. An alternative is to use the geometric mean of the prior and the current posterior approximation as the proposal density, inspired by optimal proposal schemes for sequential Approximate Bayesian Computation \citep{Alsing2018optimal}; this has the nice property that it better samples the tails of the distribution, and may hence give more robust convergence there (we find this to be the case in experiments).
\subsubsection{Active learning with Bayesian optimization}
A second approach is to use Bayesian-optimization style acquisition rules for running new simulations \citep{Lueckmann2018}. In this scheme, the next simulation is run at the parameters that maximize some deterministic acquisition function $A(\btheta)$, that encodes a trade-off between relevance (ie., being in a region of high posterior-density), and uncertainty in current learned posterior-density-surface. This active learning approach requires two ingredients: (1) some way of quantifying uncertainty in the current learned posterior surface, and (2) some carefully chosen acquisition rule. 

Training an ensemble of NDEs (as described in \S \ref{sec:ensemble}) provides a straightforward estimate of the variance of the learned likelihood surface (the weighted variance of the NDEs in the ensemble), as do Bayesian networks (the variance of the network output under the network weight posterior). Defining an optimal acquisition rule poses a more challenging problem. A simple, pragmatic acquisition rule is just the current variance of the estimated posterior density \citep{Lueckmann2018}. This has the appeal that it is cheap and simple to compute, but has the disadvantage that it does not attempt to quantify the expected improvement in the density estimator after running a new simulation in any principled way. \citet{Jarvenpaa2018} try to address this by taking a more formal decision theoretic approach: minimizing the expected integrated variance of the approximate posterior, under a new simulation draw. Whilst this is clearly a better-motivated acquisition rule, it can be computationally cumbersome in practice since it involves optimizing over high-dimensional (parameter-space) integrals.

One might expect well-chosen deterministic acquisition rules (Bayesian optimization) to be more optimal than stochastically drawing new simulation parameters from an adaptive proposal (SNL). However, \citet{Durkan2018} reported that the two approaches gave broadly similar performance in a number of experiments in the context of DELFI. This is an ongoing active area of research.
\subsection{Global versus local emulators}
For parameter inference tasks where the data have been observed in advance of the analysis, the target is the likelihood function $p(\data_\mathrm{obs} | \btheta)$. In this situation, active learning helps us to selectively run simulations in the most relevant parts of parameter space to learn the target accurately. However, in some scenarios we may run many ``experiments'' that generate independent realizations of data $\data$ from the same data generating process, and we want to analyze those data as they are taken. In these situations, it is desirable to abandon active learning and build a global emulator for $p(\data | \btheta)$ over the full prior volume, that can then be used to analyze any subsequent data $\data$ as they are observed. An example of this situation could be event reconstruction for dark matter direct detection (eg., \citealp{Simola2018}): every time an event occurs it generates some data $\data$ (the response of the detectors to the event), and we want to infer the characteristics of the event (position, energy, etc) from those data. Having a pre-trained global emulator for $p(\data | \btheta)$ would allow posterior inference from any event data $\data$ to be obtained rapidly, as the events are observed. DELFI provides a natural framework for building global emulators for data sampling distributions for these scenarios.
\section{Data compression}
\label{sec:compression}
Whether data compression is required for performing DELFI or not depends critically on the the size of the data vector relative to the number of simulations that can be feasibly performed, given that one needs enough simulations to learn the sampling distribution of the data (summaries) as a function of the parameters, over the relevant parameter-space volume. For problems with modest dimensional data-vectors, or larger data vectors but where simulation is cheap, DELFI may be performed directly on the data without further compression. However, for large datasets with expensive simulations it is clearly advantageous to compress the data down to a small number of informative summary statistics, so that the density-estimation task need only be performed on the low-dimensional data-summaries. 

In this section we review two key approaches to compressing $N$ data down to $p$ summaries -- one per parameter -- whilst aiming to retain as much information about the parameters as possible: approximate score-compression, and data compression with deep neural networks.
\subsection{Approximate score-compression}
When the likelihood function is known, the score function $\t = \nabla_{\btheta}\ln\,p(\data | \btheta)$ yields compression of $N$ data down to $p$ summaries, one per parameter, such that the Fisher information of the data is preserved (provided the gradient is taken close to the true parameters, \citealp{Alsing2018, Alsing2018delfi}). For Gaussian data where the model depends on the parameters either through the mean or the covariance, score-compression is equivalent to \textsc{moped} \citep{Heavens2000a} or the optimal quadratic estimator \citep{Tegmark1997}, respectively.

For likelihood-free applications, the likelihood-function is obviously not known a priori, but the idea of score-compression can still provide a guiding hand for defining compressed data summaries. For many problems, whilst an exact likelihood is not known, an approximate (eg. Gaussian) likelihood may still be used for defining approximate score-compressed summaries, the only cost of the approximation being some loss of information. If no obvious likelihood approximation presents itself and the data space is not too large, one can learn the conditional density $p(\data | \btheta)$ from simulations in the neighborhood of some fiducial parameters $\btheta_*$ (with an NDE), and use that to define an approximate score function. As a third approach, the score-function may be regressed directly from simulations in a likelihood-free manner using neural networks \citep{Brehmer2018mining, Brehmer2018constraining, Brehmer2018guide}.
\subsubsection{Nuisance hardened approximate score-compression}
\label{sec:hardened}
For problems with $p$ interesting parameters $\btheta$ and $m$ additional nuisance parameters $\boldsymbol\eta$, \citet{Alsing2018nuisance} (building on \citealp{Zablocki2016}) showed that it is possible to find $n$ ``nuisance hardened" score-compressed summaries (one per interesting parameter) that are insensitive-by-design to the nuisance parameters. This can be achieved through a simple projection involving the Fisher matrix \citep{Zablocki2016,Alsing2018nuisance},
\begin{align}
\label{projection}
\bar{\t}_{\btheta} =  \t_{\btheta} - \mathbf{F}_{\btheta\boldsymbol{\eta}}\mathbf{F}^{-1}_{\boldsymbol{\eta}\boldsymbol{\eta}}\t_{\boldsymbol{\eta}},
\end{align}
where $\t_{\btheta} = \nabla_{\btheta}\ln\,p(\data | \btheta)$, $\t_{\boldsymbol{\eta}} = \nabla_{\boldsymbol{\eta}}\ln\,p(\data | \btheta)$, the Fisher information matrix is given by $\mathbf{F} = -\langle\nabla_{(\btheta,\boldsymbol{\eta})}\nabla_{(\btheta,\boldsymbol{\eta})}^T\ln\,p(\data | \btheta)\rangle$, and $\bar{\t}_{\btheta}\in\mathbb{R}^p$ are the nuisance hardened summaries.

In the context of likelihood-free inference, because the score-compression and hence the nuisance parameter projection is approximate, the nuisance parameters should still be varied in the forward simulations to correctly capture any residual nuisance-sensitivity of the hardened summaries and give self-consistent nuisance marginalized posteriors (see \citealp{Alsing2018nuisance} for details). 

The ability to project nuisance parameters in this way has profound implications for likelihood-free cosmology: the complexity of the inference task (and hence number of simulations required) now only depends on the number of interesting parameters\footnote{Since DELFI involves learning $p(\t | \btheta)$, and when using nuisance hardened summary statistics, $\t\in\mathbb{R}^n$ and $\btheta\in\mathbb{R}^n$ irrespective of the presence or number of additional nuisance parameters in the problem.}, which for cosmological applications is typically relatively small ($\lesssim 10$).
\subsection{Deep neural network data compression and information maximizing networks}
An emerging trend in cosmology is to find cosmological parameter estimators from complex data sets by training deep neural networks to regress parameters from data simulations \citep{Ravanbakhsh2016,Gupta2018,Ribli2018,Fluri2018,Gillet2018}. The resulting trained networks can be viewed as radical data compression schemes, summarizing large data sets down to a set of parameter estimators whose sampling distributions (and hence likelihood functions) are unknown. These neural network parameter estimators can be straightforwardly used as data summaries in a subsequent likelihood-free analysis. However, they typically require a large number of simulations spanning the full (relevant) parameter volume in order to train.

Combining the ideas of deep network and score compression, Information Maximizing Neural Networks (IMNN; \citealp{Charnock2018}) parameterize the data compression function $\t(\data):\mathbb{R}^N\rightarrow\mathbb{R}^p$ as a neural network, training the network on a set of forward simulations such that the retained Fisher information content of the compressed summaries is maximized (see \citealp{Charnock2018} for details). This tends towards optimal non-linear compression when provided with a sufficiently flexible architecture and representative simulations\footnote{Asymptotic optimality is only expected for unimodal likelihoods and taking an expansion point close to the maximum-likelihood; this can be iterated if required.}. IMNNs have some advantages over other deep network parameter estimators. Firstly, they take fewer simulations to train, only requiring simulations around some fiducial parameters rather than spanning the full parameter volume. Secondly, by construction they implicitly (attempt to) Gaussianize the compressed summaries (and provide pseudo-maximum likelihood estimators from the transformed likelihood). This means that in a subsequent DELFI analysis, a relatively simple (close to Gaussian) conditional density estimator may be used, requiring fewer simulations to converge. Thirdly, IMNNs also provide an estimated Fisher matrix that is useful for initializing density estimators for DELFI (see \S \ref{sec:pretraining}), and also for projecting out any nuisance parameters in the compression using Eq. \eqref{projection}.

Other novel deep network compression schemes train networks to find data summaries based on their ability to distinguish (via classification) between different models \citep{Merten2018}. This is an active area of research.
\subsection{Considerations for cosmological data analysis: two-step data compression}
It is standard practice in cosmology to compress large data sets down to some set of informative summary statistics, motivated by knowledge of the underlying physics of the problem. For example, surveys are often compressed down to power spectra or higher order $n$-point statistics, supernovae lightcurves and spectra are compressed down to point estimates for their apparent magnitudes and redshifts, etc. Whilst massive data compression using the score or deep networks is, in principle, possible at the level of the raw data, we anticipate that applications of likelihood-free inference in cosmology will typically first construct a number of ``first level summaries" (eg., the usual $n$-point statistics etc), and then perform a second massive compression step on those summaries \citep{Alsing2018delfi}.

Whilst this initial compression to some first-level summary statistics may seem unnecessary (and potentially lossy), it comes with some advantages over massively compressing maps (or raw data) directly. Even sophisticated simulations may represent incomplete descriptions of the true generative data model, limited by computational resources, incomplete knowledge of the instrument or hard-to-simulate non-linear physics, etc. The first level compression step allows us to use only aspects of the data that we expect to be well-modelled by the approximate simulations. For example, problems involving an $N$-body step might resort to approximations such as \textsc{cola} \citep{Tassev2013}, which are only accurate for certain statistics and on certain scales. For cosmic microwave background analyses, noise simulations are typically expensive and approximations may be employed; cross-correlations between detector frequencies may be well-modelled, whereas the auto-power spectra may be less reliable.

On the other hand, applying flexible deep network compression schemes to the raw data offers the opportunity to learn highly informative data summaries that are not captured by standard cosmological estimators. However, this comes with the risk of learning features/approximations in the simulations that do not well-describe the data, and should therefore only be used (cautiously) for very high-fidelity simulators.
\section{pydelfi: a public code for density estimation likelihood-free inference}
\label{sec:implementation}
In this section we introduce \textsc{pydelfi} -- a flexible public code for performing density-estimation likelihood-free inference with NDEs and active learning. We briefly outline some of the implementation details and key features of the code here, referring the reader to \url{https://github.com/justinalsing/pydelfi} for tutorials and documentation.
\subsection{Overview}
Performing density-estimation likelihood-free inference with \textsc{pydelfi} proceeds as follows:

(1) Specify the architectures -- number of layers, hidden units, and activation functions -- for an ensemble of neural density estimators (MDNs, MAFs or a combination of the two).

(2) Specify a \texttt{simulator()} function that takes in parameters and returns a simulated data vector.

(3) If data compression is required, specify a \texttt{compressor()} function that takes in a data vector and returns a vector of compressed summaries. This could be an implementation of approximate-score compression, a trained deep network parameter estimator, information-maximizing network, or otherwise.

(4) Provide the observed data vector and run \textsc{pydelfi} using either the sequential neural likelihood or Bayesian optimization active learning methods, to learn the likelihood function. These are implemented as described in Algorithms \ref{algo:snl} and \ref{algo:bo} respectively. Simulation batches are run in parallel with MPI as standard, and the user has control over the number of simulations to run per round, the number of rounds to run, and the network training scheme (see below). The result is a callable likelihood function that improves during each round of new simulations and network training.

Alternatively, if a suite of simulations has been run beforehand (spanning the relevant parameter volume), these can be fed straight into \textsc{pydelfi} and the ensemble of NDEs is trained on those (without exploiting the active learning strategies). For more optimal use of resources, however, it is advantageous to provide \textsc{pydelfi} with a callable simulator so that it can exploit active learning to decide where to run simulations on-the-fly.

In the following sections we give brief details of the neural network implementation, initialization and training schemes (\S \ref{sec:pydelfi_networks}), active learning strategies (\S \ref{sec:pydelfi_active_learning}) and data compression options (\S \ref{sec:pydelfi_compression}).
\begin{algorithm}
\caption{Schematic outline of DELFI with the sequential neural likelihood method. In the description below, $\tilde p$ represents the current posterior approximation, $\pi$ denotes the prior, and $\t_o$ denotes the observed data summaries.}
\begin{algorithmic}
\State{// \texttt{Create ensemble of NDEs:}}
\State{NDEs = NDEs(\emph{chosen network architectures})}
\vspace{3mm}
\State{// \texttt{(fisher pre-training happens here if desired)}}
\vspace{3mm}
\State{// \texttt{Choose initial proposal density $q^{(0)}(\btheta)$:}}
\State{$q^{(0)}(\btheta) = $ \emph{chosen initial proposal}}
\vspace{3mm}
\State{// \texttt{SNL: run sims in batches with adaptive proposal}}
\For{$n$ in $0:n_\mathrm{rounds}$}
\For{$i$ in $0:n_\mathrm{batch}$}
\State{$\btheta_i \leftarrow q^{(n)}(\btheta)$}
\State{$\data_i \leftarrow \mathrm{simulator}(\data | \btheta_i)$}
\State{$\t_i = \t(\data_i)$}
\State{$\{\t, \btheta\}_\mathrm{training}  \leftarrow \t_i,\,\btheta_i$}
\EndFor
\State{// \texttt{train NDEs, update proposal after each round}}
\State{$\mathbf{train}$(NDEs, $\{\t, \btheta\}_\mathrm{training}$)}
\State{$q^{(n+1)}(\btheta) = \sqrt{\tilde{p}(\btheta | \t_o)\pi(\btheta)} $}
\EndFor
\vspace{3mm}
\end{algorithmic}
\label{algo:snl}
\end{algorithm}
\begin{algorithm}
\caption{Schematic outline of DELFI with Bayesian optimization. In the description below, $\tilde p$ represents the current posterior approximation, and $A$ denotes the acquisition function, and $\t_o$ denotes the observed data summaries.}
\begin{algorithmic}
\State{// \texttt{Create ensemble of NDEs:}}
\State{NDEs = NDEs(\emph{chosen network architectures})}
\vspace{3mm}
\State{// \texttt{Choose acquisition rule $A(\btheta | \mathrm{NDEs})$:}}
\State{$A(\btheta | \mathrm{NDEs})$ = \emph{chosen acquisition rule}}
\vspace{3mm}
\State{// \texttt{(fisher pre-training happens here if desired)}}
\vspace{3mm}
\State{// \texttt{run initial batch of sims with proposal} $q^{(0)}(\btheta)$}
\For{$i$ in $1:n_\mathrm{initial}$}
\State{$\btheta_i \sim q^{(0)}(\btheta)$}
\State{$\data_i \sim \mathrm{simulator}(\data | \btheta_i)$}
\State{$\t_i = \t(\data_i)$}
\State{$\{\t, \btheta\}_\mathrm{training} \leftarrow \t_i,\,\btheta_i$}
\vspace{3mm}
\EndFor
\State{// \texttt{train the NDEs}}
\State{$\mathbf{train}$(NDEs, $\{\t, \btheta\}_\mathrm{training} $)}
\vspace{3mm}
\State{// \texttt{Bayesian optimization acquisition rounds: run sims in batches at the optimal acquisition point}}
\For{$n$ in $1:n_\mathrm{rounds}$}
\State{$\btheta_n = \mathrm{argmax}\;A(\btheta | \mathrm{NDEs})$}
\For{$i$ in $1:n_\mathrm{batch}$}
\State{$\data_i \leftarrow \mathrm{simulator}(\data | \btheta_n)$}
\State{$\t_i = \t(\data_i)$}
\State{$\{\t, \btheta\}_\mathrm{training}  \leftarrow \t_i,\,\btheta_n$}
\EndFor
\State{// \texttt{train NDEs}}
\State{$\mathbf{train}$(NDEs, $\{\t, \btheta\}_\mathrm{training}$)}
\EndFor
\vspace{3mm}
\end{algorithmic}
\label{algo:bo}
\end{algorithm}
\subsection{Neural network implementation and training}
\label{sec:pydelfi_networks}
\subsubsection{Training and mitigating over-fitting}
All neural networks are implemented in \textsc{tensorflow} \citep{tensorflow2015-whitepaper}. As a default, we train the neural networks using the stochastic gradient optimizer \textsc{adam} \citep{kingma2014adam}, with a default batch-size of one tenth of the training set at each training cycle and a learning rate of $0.001$. Over-fitting is mitigated using early-stopping; during each training cycle, some fraction of the training set is set aside for validation (default 10\%), and training is terminated when the validation-loss does not improve after some user-specified threshold number of epochs (default 20). Learning rates, cross-validation fractions and early-stopping thresholds can be easily controlled by the user. The use of ensembles of networks provides additional protection against over-fitting.
\subsubsection{Ensembles and stacking}
The learned likelihood function is constructed by stacking the NDEs in the ensemble, trained with early-stopping to avoid over-fitting, weighted by their relative cross-validation losses.
\subsubsection{Initialization: Fisher pre-training}
\label{sec:pretraining}
The weights of the neural density estimators are randomly initialized by default. However, it is advantageous to exploit any expectations we might have about the sampling distribution of the data summaries to provide educated starting points for the NDEs that can then converge more quickly to the target. When using approximate-score (or IMNN) compression, the compressed summaries can be cast into pesudo maximum-likelihood estimators through a simple shift and re-scaling \citep{Alsing2018}:
\begin{align}
\label{pmle}
    \t \rightarrow \btheta_* + \mathbf{F}^{-1}_*\t.
\end{align}
This hints at a natural first guess for their sampling distribution: Gaussian estimators for the parameters, with covariance $\mathbf{F}^{-1}$. Before running any simulations, one can regress the NDEs to $p(\t | \btheta) = \mathcal{N}(\t | \btheta, \mathbf{F}^{-1})$. This initializes the NDEs to a rough (linear Gaussian) approximation of the target density, that can subsequently morph quickly towards the target when fed even a small number of simulations. We call this initialization scheme \emph{Fisher pre-training}. This is outlined in Algorithm \ref{algo:fisher} (assuming score or IMNN compressed summaries have been cast as pseudo-MLEs). As a default we draw $10^6$ pre-training data from $\mathcal{N}(\t | \btheta, \mathbf{F}^{-1})$, with parameters drawn from the prior. The pre-training data are discarded after the initialization of the NDEs.

In a similar spirit, the NDEs can alternatively be initialized by running DELFI on any cheap approximate simulations that might be available, and then discarding the approximate sims when training on full simulations. 

If no good approximate starting point or approximate simulations are available, the network weights are initialized randomly by default.
\subsection{Active learning}
\label{sec:pydelfi_active_learning}
\subsubsection{Sequential neural likelihood}
We implement SNL as outlined in Algorithm \ref{algo:snl}. The user can specify an initial proposal for running the first batch of simulations. After each round of training, we draw new parameters for simulating from an updated proposal, $q(\btheta) = \sqrt{\tilde{p}(\btheta | \t_o)\pi(\btheta)}$ -- the geometric mean of the prior $\pi$ and the current posterior approximation $\tilde p $ (inspired by \citealp{Alsing2018optimal}).

During each simulation acquisition batch, simulations are run in parallel with MPI. The number of simulations per batch is chosen by the user.
\subsubsection{Bayesian optimization}
We implement active learning with Bayesian optimization as described in Algorithm \ref{algo:bo}. The implemented acquisition function is the estimated posterior variance, calculated from the ensemble of NDEs; bespoke acquisition functions can be implemented by the user if needed.

Whilst simulations can be acquired one-by-one in this set-up, where parallel computing is available it is typically desirable to run many simulations concurrently. Therefore, \textsc{pydelfi} runs batches of simulations in parallel (with MPI) at each derived acquisition point as default.
\subsection{Data compression}
\label{sec:pydelfi_compression}
\textsc{pydelfi} comes with classes for approximate-score compression for common exponential family data-distributions. Where expectation values, covariances, derivatives, etc., need to be estimated from forward simulations, these are run in parallel with MPI as standard, otherwise pre-computed or analytical approximations can be fed in. For IMNN compression, a public implementation is available at \url{https://github.com/tomcharnock/IMNN}.

\textsc{pydelfi} has the flexibility to take any (or no) compression scheme; bespoke compression schemes can be straightforwardly defined by the user and fed into \textsc{pydelfi}.
\begin{algorithm}
\caption{Schematic outline of the Fisher pre-training step described in \S \ref{sec:pretraining}. $\mathbf{F}$ denotes the approximate Fisher matrix, $q^\mathrm{pre-training}$ the parameter-proposal distribution for the pre-training data (taken to be the prior, by default).}
\begin{algorithmic}
\State{// \texttt{fisher pre-training:}}
\vspace{3mm}
\State{// \texttt{generate pre-training data}}
\For{$i$ in $1:n_\mathrm{pre-training}$}
\State{$\btheta_i\sim q^\mathrm{pre-training}(\btheta)$}
\State{$\t_i\sim \mathcal{N}(\t | \btheta, \mathrm{F}^{-1})$}
\State{$\{\t,\btheta\}_\mathrm{pre-training}\leftarrow \t_i,\,\btheta_i$}
\EndFor
\vspace{3mm}
\State{// \texttt{train the NDEs}}
\State{$\mathbf{train}$(NDEs, $\{\t, \btheta\}_\mathrm{pre-training}$)}
\end{algorithmic}
\label{algo:fisher}
\end{algorithm}
\section{Case study I (validation): JLA Supernovae analysis}
\label{sec:jla}

Supernova data analysis is an interesting opportunity for likelihood-free methods, since the data are impacted by a large number of systematic biases and selection effects that need to be carefully accounted for to obtain robust cosmological parameters. Whilst progress has been made recently in developing Bayesian hierarchical models (BHMs) that attempt to carefully treat these effects \citep{Mandel2009, March2011, Rubin2015, Shariff2016, Roberts2017, Hinton2018}, likelihood-free methods have the advantage that the forward model complexity is unfettered by practical limitations of implementing and sampling high-dimensional BHMs.

As a validation test, we perform a simple analysis of the JLA data \citep{Betoule2014} under assumptions that allow us to compare against an exact known likelihood. The set-up is identical to \citet{Alsing2018delfi}, which we review briefly below.
\subsection{JLA data and model}
\begin{figure*}
\centering
\includegraphics[width = 16cm]{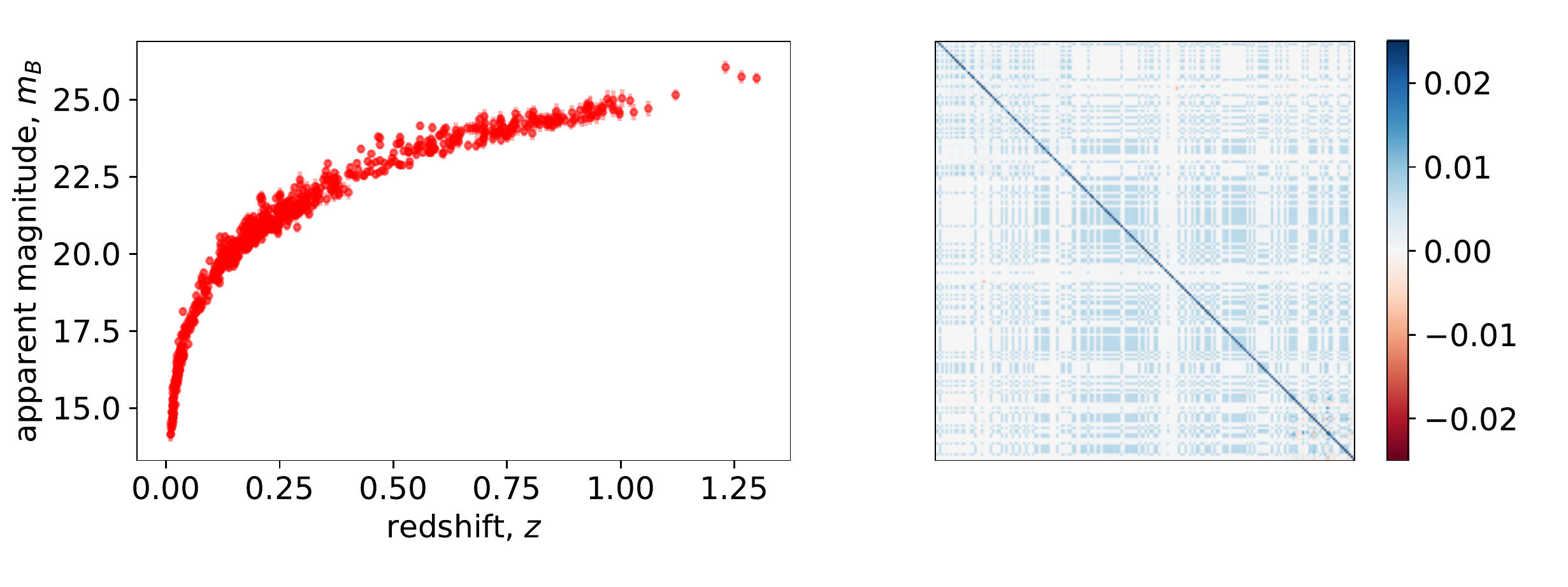}
\caption{Left: Measured apparent magnitudes (with uncertainties) against their measured redshifts for the JLA supernova sample. Right: Covariance matrix corresponding to the observed apparent magnitudes.}
\label{fig:jla_data}
\end{figure*}

The JLA sample is comprised of $740$ type Ia supernovae with estimated apparent magnitudes $m_\mathrm{B}$, redshifts $z$, color at maximum-brightness $C$ and stretch $X_1$ parameters. We take the data vector to be the vector of estimated apparent magnitudes $\data = (\hat{m}_\mathrm{B}^1, \hat{m}_\mathrm{B}^2,\dots,\hat{m}_\mathrm{B}^M)$, where uncertainties in $z$, $C$ and $X_1$ are implicitly accounted for in the covariance matrix (see \citealp{Betoule2014}, also Figure \ref{fig:jla_data}).

We take the expected apparent magnitudes of type Ia supernovae to be given by \citep{Tripp1998},
\begin{align}
\label{apparent_mag}
m_\mathrm{B} = 5\mathrm{log}_{10}\left[\f{D^*_\mathrm{L}(z ; \btheta)}{10\mathrm{pc}}\right] &- \alpha X_1 + \beta C \nonumber \\
& + M_\mathrm{B} + \delta M\,\Theta(M_\mathrm{stellar} - 10^{10}M_\odot)
\end{align}
where $D^*_\mathrm{L}$ is the luminosity distance (at reference $h=1$), $\btheta$ are the cosmological parameters (see below), $\alpha$ and $\beta$ are calibration parameters for the stretch and color, and $M_\mathrm{B}$ and $\delta M$ characterize the host stellar-mass dependent reference absolute magnitude. $\Theta$ is the Heaviside function.

We assume a flat $w$CDM cosmology parameterized by matter density $\Omega_\mathrm{m}$ and dark energy equation-of-state $p/\rho=w_0$.
\subsection{Simulations}
For this validation case, simulations are just draws from the (exact) Gaussian sampling distribution of the data, ie., drawing Gaussian data from Eq. \eqref{jla_sampling}.
%
%
%
%
\begin{figure*}
\centering
\includegraphics[width = 17.5cm]{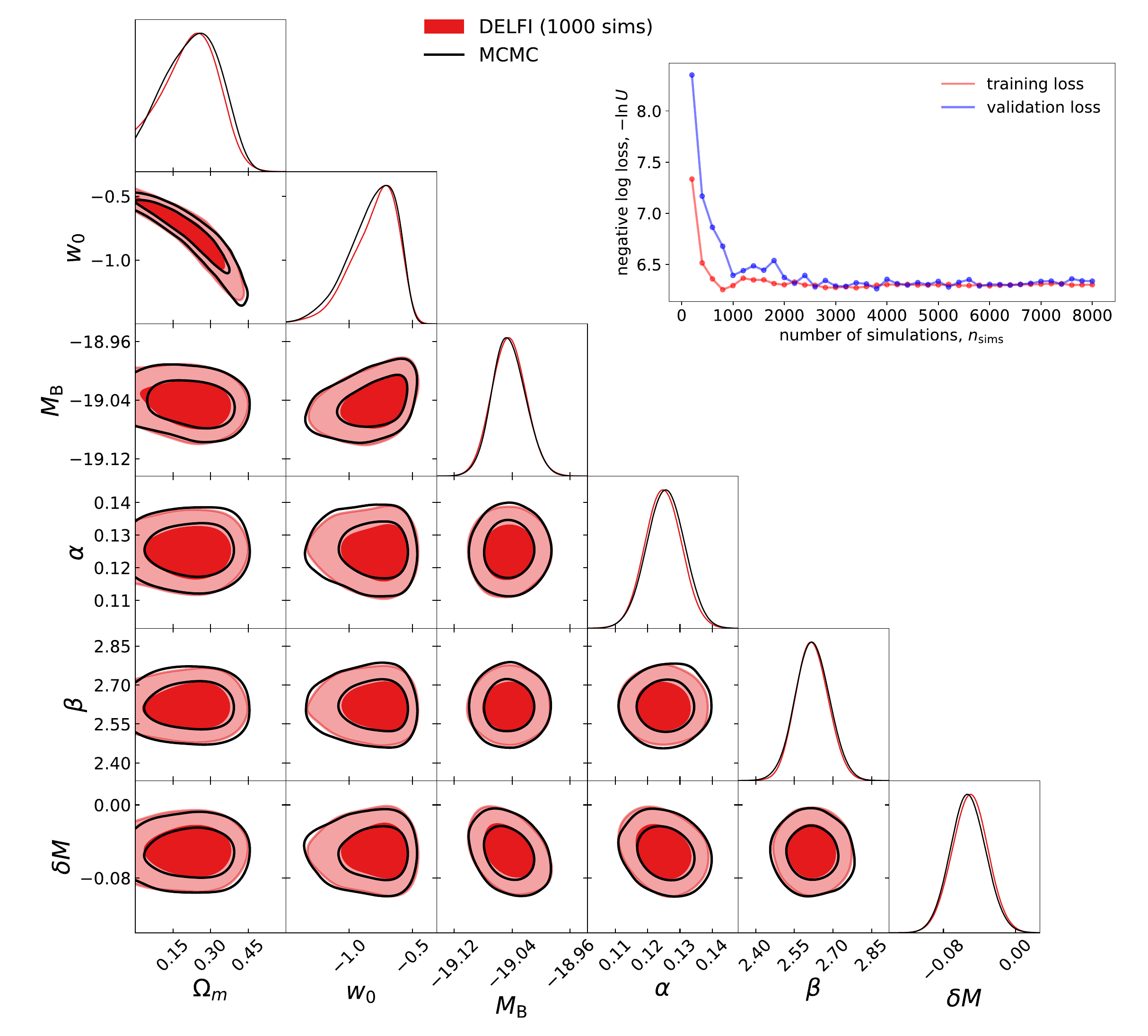}
\caption{68 and 95\% credible regions of the 2D projections of the inferred posteriors from a long-run MCMC chain (black) and DELFI from $1000$ forward simulations (red). The posteriors from DELFI (after just 1000 simulations) and the long-run MCMC chain are in excellent agreement.}
\label{fig:jla_contours}
\end{figure*}
\subsection{Data compression}
For this validation case we assume the data are Gaussian, 
\begin{align}
\label{jla_sampling}
\ln\,p(\data | \bphi) = -\f{1}{2}(\data - \boldsymbol\mu(\bphi))^T\cov^{-1}(\data - \boldsymbol\mu(\bphi)) - \f{1}{2}\ln |\cov|,
\end{align}
with mean given by Eq. \eqref{apparent_mag}, and we assume a fixed covariance matrix from \citet{Betoule2014} (see also \citealp{Alsing2018delfi} for details of the covariance matrix).

For data compression, we use the score of the Gaussian likelihood:
\begin{align}
\t \equiv\nabla_{\btheta}\mathcal{L}_*= \nabla_{\btheta}^T\boldsymbol\mu_*\cov^{-1}(\data - \boldsymbol\mu_*),
\end{align}
where `$*$' indicates evaluation at fiducial parameters $\btheta_* = (0.202, -0.748, -19.04, 0.126, 2.644, -0.0525)$\footnote{Found in a few iterations of the pseudo maximum-likelihood estimator, Eq. \eqref{pmle}.}.
\subsection{Priors}
We assume broad Gaussian priors on the parameters $\btheta= (\Omega_\mathrm{m}, w_0, \alpha, \beta, M_\mathrm{B}, \delta M)$ with mean and covariance:
\begin{align}
&\boldsymbol{\mu}_\mathrm{P} = (0.3,\;  -0.75 ,\; -19.05 ,\;   0.125,\; 2.6  ,  \;-0.05),\nonumber \\
&\cov_\mathrm{P}=
  \left( {\begin{array}{cccccc}
   0.4^2 & -0.24 & 0 & 0 & 0 & 0 \\
   -0.24 & 0.75^2 & 0 & 0 & 0 & 0 \\
   0 & 0 & 0.1^2 & 0 & 0 & 0 \\
   0 & 0 & 0 & 0.025^2 & 0 & 0 \\
   0 & 0 & 0 & 0 & 0.25^2 & 0 \\
   0 & 0 & 0 & 0 & 0 & 0.05^2 \\
\end{array} } \right),
\end{align}
with additional hard prior boundaries on $\Omega_\mathrm{m}\in[0, 0.6]$ and $w_0\in[-1.5, 0]$.
\subsection{DELFI set-up}
We ran DELFI using the SNL active learning scheme as described in \S \ref{sec:implementation}. An ensemble of six NDEs was used: five MDNs with 1--5 Gaussian components respectively, each with two hidden layers of 50 hidden units, and a MAF containing five MADEs, each with two hidden layers of 50 units. We use $\mathrm{tanh}$ activation functions throughout.

Simulations were run in batches of $250$ after an initial Fisher pre-training step to initialize the network ensemble.
\subsection{Results}
Fig. \ref{fig:jla_contours} (inset) shows the convergence of the DELFI NDE-ensemble as a function of the number of simulations. Convergence is achieved after $\mathcal{O}(10^3)$ simulations. This is a substantial improvement on the $20,000$ simulation requirement reported for the same problem in \citet{Alsing2018delfi}. This also represents a substantial improvement over the Bayesian optimization likelihood-free inference (BOLFI) approach presented in \citet{Leclercq2018}. That work reported $6000$ simulations were required for the same toy JLA analysis problem but only inferring two parameters with the other four held fixed. The BOLFI approach implemented in that work also requires much stronger assumptions about the sampling distribution of the data, implicitly assuming that the data are Gaussian with a known covariance matrix (but unknown mean).

Fig. \ref{fig:jla_contours} shows the recovered DELFI posterior after $1000$ simulations (red), against a long-run MCMC chain for validation (black). The DELFI and MCMC posteriors are in excellent agreement.
\subsection{Discussion}
This simple validation case gives insight into the relative performance of DELFI and MCMC sampling for simple problems where the sampling distribution of the data is known, and the likelihood can be evaluated exactly for given model parameters. In the JLA example above, DELFI was well converged after $\mathcal{O}(10^3)$ forward simulations (which are of the same cost as likelihood evaluation for this case), while MCMC sampling typically requires one or two orders-of-magnitude more likelihood calls to give well-converged chains. Since training neural density estimators on small training sets is computationally inexpensive, in the many cases where the cost of DELFI is dominated by making draws from $p(\data | \btheta)$ DELFI offers a fast and accurate alternative to MCMC sampling, giving orders of magnitude speed-up for typical $\sim 6$ parameter problems. The number of simulations required for DELFI to converge for these known-likelihood problems can be further reduced by training an NDE that corresponds to the known data sampling distribution, ie., a Gaussian with parameter-dependent mean and fixed covariance matrix for this case.

The advantage of DELFI over MCMC for these known-likelihood inference problems is due in part to the data compression step, turning the inference task into a low-dimensional conditional density estimation task. For unimodal likelihoods, the score provides asymptotically optimal compressed summaries and is readily available, and application of DELFI as a replacement for MCMC is straightforward. However, for multimodal likelihoods, more care must be taken in defining approximately sufficient statistics for the problem at hand.
\section{Case study II: tomographic cosmic shear pseudo-\texorpdfstring{$C_\ell$}{Cl} analysis}
\label{sec:cosmic_shear}
Cosmic shear data are well suited to likelihood-free analyses, containing a large number of effects that may be simulated (to varying degrees) but are challenging to build into an accurate likelihood function. Non-linear physics and baryonic feedback \citep{Rudd2008,Harnois2015}, intrinsic alignments \citep{Joachimi2015galaxy}, shape and photo-$z$ measurement systematics \citep{Massey2012,Mandelbaum2018,Salvato2018}, image blending \citep{Mandelbaum2018}, reduced shear corrections \citep{Krause2010}, non-trivial non-Gaussian sampling distributions for common summary statistics \citep{Sellentin2018}, the redshift-dependent source galaxy population model \citep{Kannawadi2018}, etc., all have the potential to bias parameter inferences if not carefully accounted for. As well as promising more principled inference, LFI may also open up the possibility to extract extra information from non-standard lensing observables (eg., magnification \citealp{Hildebrandt2009,VanWaerbeke2010, Hildebrandt2013, Duncan:2013bb, Heavens2013, Alsing2015}) and non-linear scales via, eg., peak counts \citep{Kratochvil2010,Fluri2018peaks}, bispectrum \citep{Cooray2001} etc.

For this simple demonstration, we perform cosmological parameter inference from tomographic shear pseudo-$C_\ell$s for a Euclid-like survey. We focus on the large-scales where the pseudo-$C_\ell$ likelihood is intractable and standard Gaussian likelihood approximations are expected to break down.
\subsection{Tomographic shear data and model}
As light from distant galaxies propagates through the Universe on its way to us, it gets gravitationally lensed by the intervening large-scale structure, imprinting a coherent distortion on the galaxy images observed on the sky. This coherent lensing distortion field provides a unique probe of both the evolution of the 3D matter distribution, and geometry of the Universe via the distance-redshift relation. In particular, the observed shapes of galaxies are modified by the lensing ``cosmic shear" fields, with their ellipticities $\epsilon$ picking up an additive distortion (in the weak lensing limit):
\begin{align}
    \epsilon = \epsilon_\mathrm{int} + \gamma,
\end{align}
where $\epsilon_\mathrm{int}$ is the unobserved intrinsic (unlensed) ellipticity, and $\gamma$ is the additional shear due to gravitational lensing. The statistical properties of the shear field $\gamma$ provide a sensitive probe of cosmology (see eg., \citealp{Kilbinger2015} for a review).

A weak lensing survey involves measuring the shapes, redshifts and angular positions on the sky of a large number of galaxies, which are then used to constrain cosmological parameters by eliciting the statistics of the cosmic shear signal in the data. We will take our data vector to be a set of (pixelized) shear maps estimated from a lensing survey $\data = (\boldsymbol\gamma^{(1)}, \boldsymbol\gamma^{(2)}, \dots, \boldsymbol\gamma^{(n_z)})$, with galaxies grouped into $n_z$ tomographic redshift bins (based on their estimated redshifts). The estimated shear in a given tomographic bin $\alpha$ and pixel $p$ is taken to be:
\begin{align}
\label{shear_pixel}
    \gamma^{(\alpha)}_p = \sum_{i\in (p,\alpha)} \hat{\epsilon}_i / N^{(\alpha)}_p,
\end{align}
ie., the average estimated ellipticity of the $N^{(\alpha)}_p$ galaxies in that pixel. The unknown intrinsic ellipticities are assumed to be zero mean random variates with standard deviation $\sigma_e$, giving Gaussian ``shape noise" $\sigma_p^{(\alpha)} = \sigma_e/\sqrt{N^{(\alpha)}}$ on each pixel (in the limit of many galaxies per pixel). The shape noise will invariably be anisotropic due to varying number of sources per pixel, and maps will be substantially masked due to incomplete sky coverage, masking around bright sources in the survey etc. 

\begin{figure*}
\centering
\includegraphics[width = 17cm]{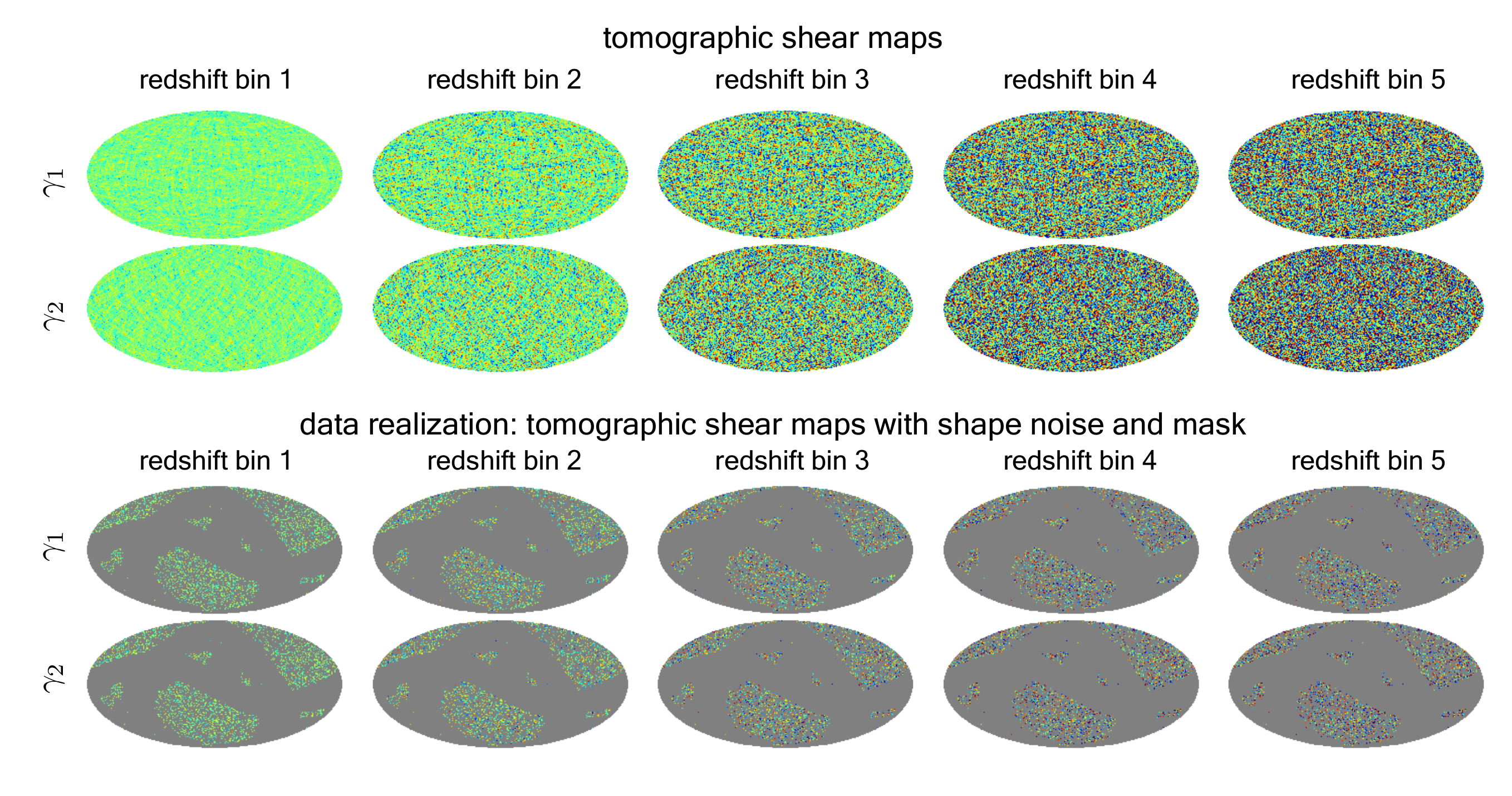}
\caption{Schematic of the mock tomographic cosmic shear data. Top: realization of Gaussian tomographic cosmic shear fields (generated for the fiducial cosmology). The two components of the complex, spin-2 shear field are shown: $\gamma\equiv \gamma_1 + i\gamma_2$. Bottom: the realized maps but with shape noise and mask added. The shape noise levels and mask are taken for a Euclid-like survey. The maps are subsequently compressed down to a small set of summary statistics in two steps: firstly, maps are compressed to auto- and cross- angular (E-mode) power spectra, and these power spectra are then further compressed using approximate-score compression (\S \ref{sec:shear_compression}).}
\label{fig:shear_data}
\end{figure*}

Mock data for this case study are generated for a survey set-up similar to the upcoming ESA \emph{Euclid} survey \citep{Laureijs2011} (as described in \S \ref{sec:cs_simulations}), and are shown in Figure \ref{fig:shear_data}.
\subsubsection{Tomographic shear power spectra}
In this case study we will focus on extracting information from the tomographic power spectra of the cosmic shear fields. 

For a given (flat) cosmological model and parameters, the predicted angular power spectra between tomographic redshift bins $\alpha$ and $\beta$ are given by\footnote{In the Limber approximation, \citep{Limber1954}.} \citep{Kaiser1992, Kaiser1998, Hu1999, Hu2002a, Takada2004, Kitching2017},
\begin{align}
\label{limber_power}
C_{\ell,\alpha\beta}^{\gamma\gamma} &= \int \f{d\chi}{\chi^2}\;w_\alpha(\chi)w_\beta(\chi)\left[1+z(\chi)\right]^2P_\delta\left(\f{\ell}{\chi}; z(\chi)\right),
\end{align}
with comoving distance-redshift relation $\chi(z)$, matter power spectrum $P_\delta(k\; \chi)$, and lensing weight functions given by
\begin{align}
\label{weight}
w_\alpha(\chi)&=\f{3\Omega_\mathrm{m}H_0^2}{2}\chi\int_{\chi}^{\chi_\mathrm{H}} d\chi'\;n_\alpha(\chi')\f{\chi'-\chi}{\chi'},
\end{align}
where $n_\alpha(\chi)d\chi = p_\alpha(z)dz$ is the redshift distribution for galaxies in redshift bin $\alpha$. Cosmological parameters enter in both the matter power spectrum and distance-redshift relation. We will assume a flat $\Lambda$CDM cosmology with parameters $\btheta = (\sigma_8, \Omega_\mathrm{m}, \Omega_\mathrm{b}, h, n_s)$.
%
%
%
\begin{figure*}
\centering
\includegraphics[width = 17.5cm]{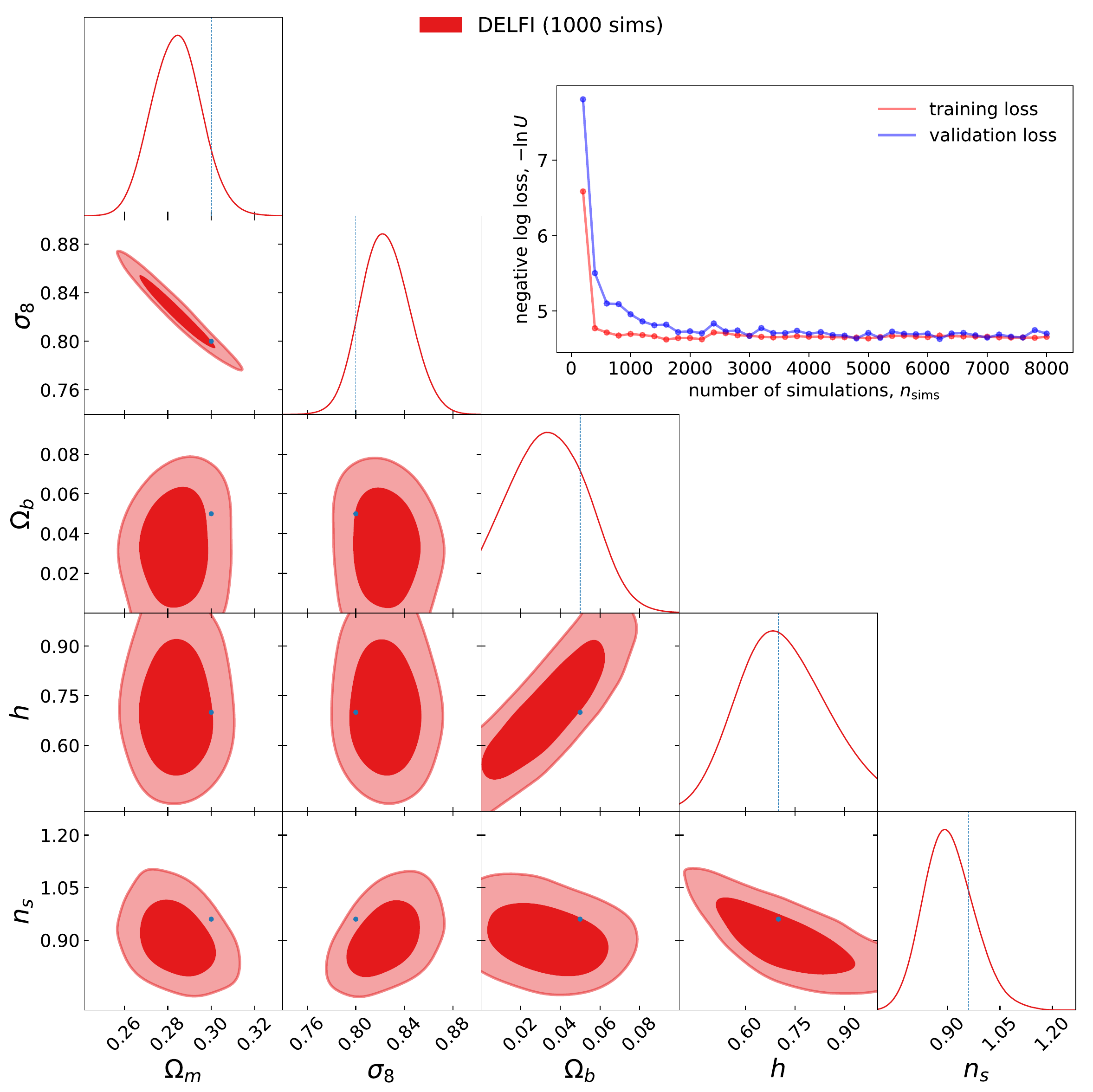}
\caption{68 and 95\% credible regions of the 2D projections of the inferred DELFI posterior after 1000 simulations, for the cosmic shear tomographic pseudo-$C_\ell$ case study. Input parameters (blue) are well recovered, within uncertainties.}
\label{fig:cs_contours}
\end{figure*}
\subsection{Simulations}
\label{sec:cs_simulations}
The simulations for this demonstration proceed as follows:
\begin{enumerate}
    \item Simulate Gaussian random tomographic shear maps (in healpix pixelization \citep{Gorski2005} with $n_\mathrm{side}=128$, $\ell_\mathrm{max} = 3n_\mathrm{side}-1=383$), with power spectrum corresponding to the input cosmology $\btheta$ (cf., Eq. \eqref{limber_power}).
    \item Add (anisotropic) shape noise to the healpix maps.
    \item Apply Euclid-like mask (footprint and star-mask; Figure \ref{fig:shear_data}).
    \item Compute tomographic pseudo-$C_\ell$ auto- and cross- band powers from the noisy tomographic maps.
\end{enumerate}
We assume a survey set-up similar to the upcoming ESA \emph{Euclid} survey \citep{Laureijs2011}: $15,000$ square degrees with a mean galaxy number density of $\bar{n}=30\,\mathrm{arcmin}^{-2}$, an overall galaxy redshift distribution $n(z) \propto z^2\mathrm{exp}\left[-(1.41z/z_m)^{1.5}\right]$ with a median $z_m = 0.9$, Gaussian photo-$z$ errors with standard deviation $\sigma_z = 0.05*(1+z)$, and five tomographic bins with equal mean galaxy number density per bin. Modes are binned into ten log-spaced bands between $\ell=10$ and $\ell=383$.

For the shape noise, we add zero-mean Gaussian noise to each pixel with variance $\sigma_e^2/N^{(i)}_p$, where $\sigma_e=0.3$ and $N_p$ is the number of galaxies in pixel $p$, tomographic bin $i$. Galaxies are Poisson distributed among pixels according to the mean number density per tomographic slice to give realistic, anisotropic shape noise.

The mock data for this demonstration are simulated following the procedure above, assuming a Planck 2018 cosmology \citep{Planck2018}: $\sigma_8 = 0.811$, $\Omega_\mathrm{m} = 0.315$, $\Omega_\mathrm{b} = 0.049$, $h = 0.674$ and $n_s = 0.965$, $w_0=-1.03$.
\subsection{Data compression}
\label{sec:shear_compression}
We compress the noisy, masked tomographic shear maps down in two steps. First, we compute from the maps a set of tomographic auto- and cross- angular pseudo-$C_\ell$ band powers, for $K$ $\ell$-bands and $n_z$ tomographic bins:
\begin{align}
    \tilde{\data} = (\hat{C}_{\mathcal{B}_1, 11}, \hat{C}_{\mathcal{B}_1, 12}, \dots, \hat{C}_{\mathcal{B}_1, n_zn_z}, \hat{C}_{\mathcal{B}_2, 11},\dots,\hat{C}_{\mathcal{B}_K, n_z, n_z})
\end{align}
where
\begin{align}
    \hat C_{\mathcal{B}_k, ij} = \sum_{\ell\in\mathcal{B}_k}\sum_{m=-\ell}^{\ell} \hat{a}^{(i)}_{\ell m}\hat{a}^{(j)*}_{\ell m},
\end{align}
and $\{\hat{a}^{(i)}_{\ell m}\}$ are the $E$-mode spherical harmonic coefficients of the noisy, masked, tomographic shear maps. Note that in the likelihood-free framework, there is no need to deconvolve the mask or subtract the noise bias from the estimated band powers; these are all taken care of (exactly) in the forward simulations.

Secondly, we compress the tomographic band powers $\tilde{\data}$ assuming they are approximately Gaussian distributed (ie., \textsc{moped} compression \citealp{Heavens2000a}), giving compressed summaries:
\begin{align}
\t \equiv\nabla_{\btheta}\mathcal{L}_*= \nabla_{\btheta}^T\boldsymbol\mu_*\cov^{-1}(\tilde{\data} - \boldsymbol\mu_*),
\end{align}
taking fiducial parameters $(\sigma_8, \Omega_\mathrm{m}, \Omega_\mathrm{b}, h, n_s) = (0.8, 0.3, 0.05, 0.7, 0.96)$ for performing the compression. We estimate the covariance and mean by running $10^3$ forward simulations. The derivatives are estimated as a forward difference using $100$ pairs of simulations per parameter, with matched random seeds to suppress sample variance, and step sizes of $5\%$ for each parameter respectively. The extra simulation burden here could easily be eliminated by using analytical models for the mean (masked) band powers and covariance matrix in place of Monte Carlo estimates (we use Monte Carlo estimates here for convenience reasons only).

Note that the requirements on the accuracy of the mean, covariance and derivatives used for the data compression are much less onerous than for an approximate Gaussian likelihood-based analysis: any errors in these estimated quantities can \emph{only} lead to sub-optimality in the compression, in contrast to a likelihood-based analysis where errors/incorrect-assumptions can bias parameter inferences.

Since the pseudo-$C_\ell$s are not expected to be exactly Gaussian distributed (particularly at low $\ell$), the second compression step may lose a small amount of information.
\subsection{Priors}
We assume broad independent Gaussian priors over $\btheta = (\sigma_8, \Omega_\mathrm{m}, \Omega_\mathrm{b}, h, n_s)$ with means $\boldsymbol{\mu}_{\btheta} = (0.8, 0.3, 0.05, 0.7, 0.96)$, standard deviations $\boldsymbol{\sigma}_{\btheta} = (0.3, 0.3, 0.1, 0.3, 0.3)$, and hard parameter limits $\sigma_8 \in \left[0.4, 1.2\right]$,  $\Omega_\mathrm{m} \in \left[0, 1\right]$, $\Omega_\mathrm{b} \in \left[0, 0.3\right]$, $h \in \left[0.4, 1\right]$, and $n_\mathrm{s} \in \left[0.7, 1.3\right]$.
\subsection{DELFI set-up}
We ran DELFI using the SNL active leaning scheme described in \S \ref{sec:implementation}. An ensemble of six NDEs was used: five MDNs with 1--5 Gaussian components respectively, each with two hidden layers of 50 hidden units, and a MAF containing five MADEs, each with two hidden layers of 50 units. We use $\mathrm{tanh}$ activation functions throughout.

Simulations were run in batches of $200$ after an initial Fisher pre-training step to initialize the network ensemble.
\subsection{Results}
Figure \ref{fig:cs_contours} (inset) shows the convergence of the DELFI NDE-ensemble as a function of the number of forward simulations; convergence is achieved after $\mathcal{O}{10^3}$ forward simulations. Figure \ref{fig:cs_contours} shows that the input cosmological parameters are well recovered (within uncertainties).
\subsection{Discussion}
The forward modelling assumptions described above are the same as those used in hierarchical modelling approaches to cosmic shear parameter inference \citep{Alsing2015hierarchical, Alsing2016}; even in this simple demonstration we could make certain more principled assumptions about the statistical model for the data than standard Gaussian-likelihood cosmic shear analyses with relative ease. While the Bayesian hierarchical approach samples the likelihood of the noisy map data and infers the tomographic shear fields explicitly as a by-product, the likelihood-free approach analyzes the compressed data and implicitly marginalizes over the latent shear maps, targeting the posterior distribution of the cosmological parameters only. In this context, the likelihood-free analysis can be viewed as a fast alternative to sampling a full hierarchical model, with the caveat that some information may be lost in the data compression step(s). However, the likelihood-free framework will allow us to extend the forward model to describe the data at the catalog or image level -- complexity that would quickly become intractable for hierarchical modelling approaches.

In this simple demonstration we made the simplifying assumptions of Gaussian shear fields and considered power spectra only in the first-level compression step. This can naturally be extended to non-Gaussian lensing simulations, and higher-order statistics added to the list of first-level summaries.
\section{Case study III: Ionizing background from high-z lyman-\texorpdfstring{$\alpha$}{a} forests}
\label{sec:Lya}
\begin{figure}
\centering
\includegraphics[width = 8cm]{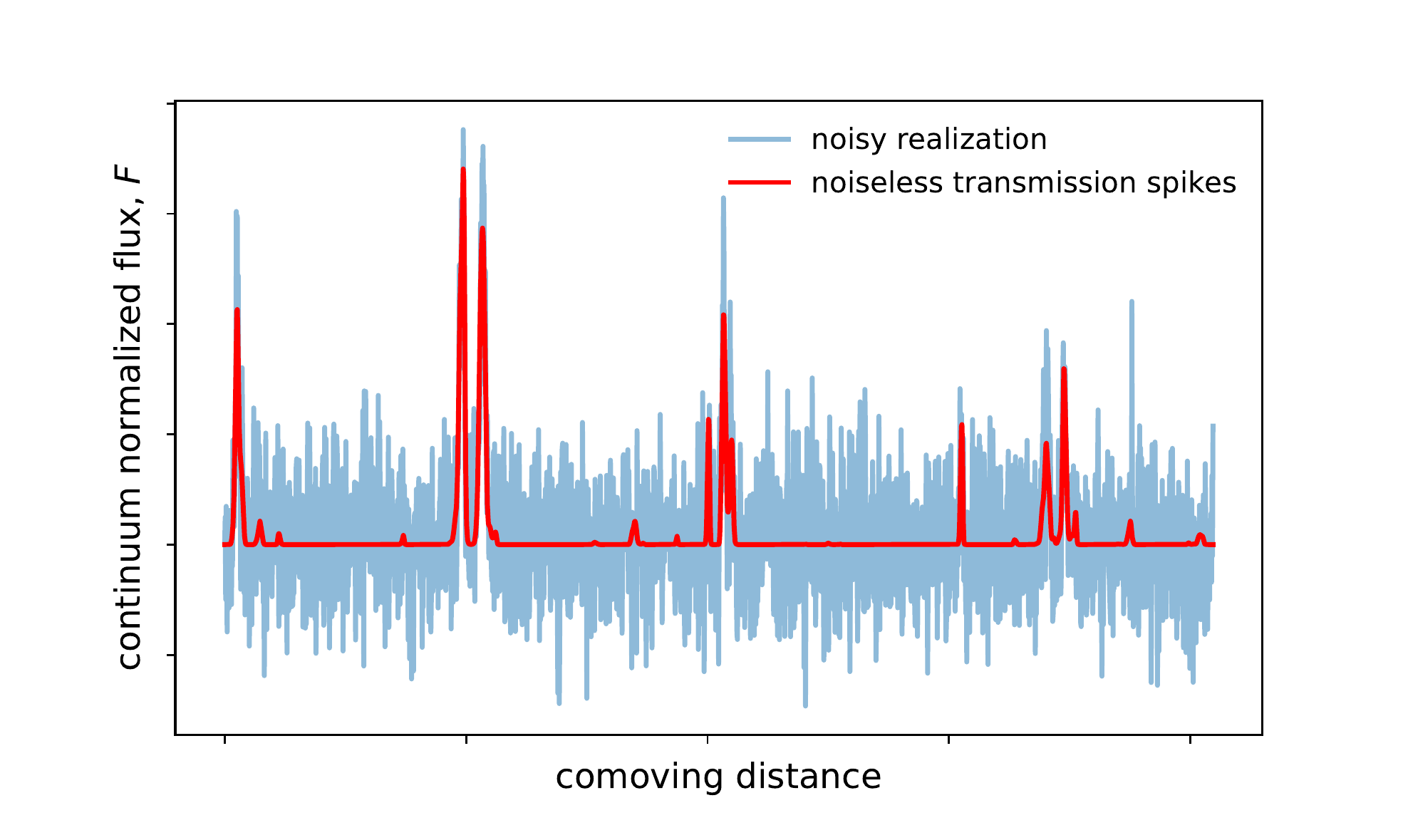}
\caption{Typical example of Ly$\alpha$ forest transmission spikes at $z\sim 6$ (red) spanning comoving distance $40 Mpc/h$, and the same forest segment with observational noise added.}
\label{fig:flux_train}
\end{figure}
The Lyman-$\alpha$ (Ly$\alpha$) forest at $z\sim 6$ measured from high redshift quasar spectra probes the ionizing background and thermal state of the intergalactic medium (IGM) around the end of the epoch of reionization (see \citealp{McQuinn2016} for a review). At these redshifts, the Universe is largely opaque to Ly$\alpha$; the forest is characterized by narrow Ly$\alpha$ transmission spikes corresponding to small, low density regions \citep{Oh2005}, separated by extended Gunn-Peterson troughs \citep{Gunn1965} where Ly$\alpha$ is completely absorbed (see Figure \ref{fig:flux_train}). The transmitted fraction of the quasar flux is given by $F\equiv e^{-\tau_{\mathrm{Ly}\alpha}}$, where the optical depth $\tau_{\mathrm{Ly}\alpha}\propto T^{-0.7}\Delta_b^2/\Gamma_{\mathrm{HI}}$ depends on the temperature $T$, gas density $\Delta_b$ and HI ionization rate $\Gamma_{\mathrm{HI}}$. The statistics of the Ly$\alpha$ transmission spikes can hence be used to constrain the ionization rate (and thermal state), but the likelihood function for the observed flux transmission is intractable; likelihood-free inference is required to draw principled inferences from these data. For a recent application of ABC in this context, see \citet{Davies2017}.

In this demonstration we will show how DELFI can be used to infer the ionization rate $\Gamma_{\mathrm{HI}}$ from observed segments of Ly$\alpha$ forest at $z\sim 6$, using hydrodynamical simulations to forward model the Ly$\alpha$ transmission. In this toy demonstration we will recover the HI ionization rate assuming a uniform ionizing background, fixed thermal state and consider Ly$\alpha$ only. However, we highlight that likelihood-free methods offer exciting new prospects for constraining poorly understood inhomogeneous reionization processes and thermal histories from high-$z$ Ly$\alpha$ and Ly$\beta$ forest observations \citep{Davies2016,dAloisio2017,Davies2017}.
\subsection{Data and simulations}
We simulate mock Ly$\alpha$ forest segments for a given ionization rate $\Gamma_\mathrm{HI}$ using the Sherwood hydrodynamical simulation suite \citep{Bolton2016}, as follows: 
\begin{enumerate}
    \item Generate a random skewer through a $z=6$ snapshot of a $40 \mathrm{Mpc}/h$ hydro-simulation box (from the Sherwood suite), ran with a fiducial ionization rate $\Gamma^*_\mathrm{HI} = 2.56\cdot10^{-13}\mathrm{s}^{-1}$ at $z = 6$ (see \citealp{Bolton2016} for details of the Sherwood simulation set-up). The HI fraction is computed in (2048) cells along the line-of-sight, assuming ionization equilibrium, and the resulting Ly$\alpha$ transmission fraction $F$ calculated (including the effects of peculiar motions and thermal broadening).
    \item The transmission flux $F$ is then re-scaled by $e^{-\Gamma^*_\mathrm{HI}/\Gamma_\mathrm{HI}}$ to impose the ionization rate we want to simulate. Note that in this simple demonstration we are fixing the instantaneous temperature and thermal history to their default values from the Sherwood suite, and also neglect large-scale fluctuations in $\Gamma_\mathrm{HI}$ that are expected to arise from inhomogeneous reionization.
    \item Add zero mean Gaussian noise to the flux values, with standard deviation $\sigma = 0.01$.
\end{enumerate} 
Mock data are generated from the above simulation pipeline with a fiducial ionization rate $\Gamma^*_\mathrm{HI} = 2.56\cdot10^{-13}\mathrm{s}^{-1}$, show in Figure \ref{fig:flux_train}. The same pipeline is then used to generate forward simulations for inferring $\Gamma_\mathrm{HI}$ from those data using DELFI.
\subsection{Data compression}
In this simple demonstration we compress the flux data-vector in two stages: First, we compute fifty percentiles of the $2048$ flux values, from 2 to 100 in steps of 2\%. This is motivated by the notion that the most of the information about the ionization rate should be contained in the PDF of the flux values, which can be conveniently summarized by a set of percentiles. 

We then compress the vector of percentiles down to a single summary statistic for $\Gamma_\mathrm{HI}$ using an IMNN. We use a fully-connected network with three dense layers with $128$, $64$ and $32$ hidden units respectively, and leaky-ReLu activation functions with activation parameter $\alpha_\mathrm{ReLu} = 0.01$. For the training set we use $5000$ simulations at the fiducial $\Gamma_\mathrm{HI}^*$, and an additional $5000$ random-seed matched simulation pairs with $\Gamma_\mathrm{HI} = \Gamma_\mathrm{HI}^*\pm 1\cdot 10^{-13}$ for the derivatives\footnote{Note that given a hydrosimulation box, generating realizations of Ly$\alpha$ segments by taking skewers through the box is inexpensive. We therefore made no attempt to reduce/optimize the number of simulations needed to train the IMNN in this case study. We leave detailed exploration of optimal compression of Ly$\alpha$ forsets (eg., without pre-compression to percentiles, optimal IMNN architectures, etc) to future work.}.
\subsection{Priors}
We take a uniform prior $\Gamma_\mathrm{HI}\in\left[0, 6\cdot10^{-13}\right]\,\mathrm{s}^{-1}$.
\subsection{DELFI set-up}
We ran DELFI using the SNL active learning scheme. We use an ensemble of five neural density estimators: five MDNs with 1--5 Gaussian components respectively, each with two hidden layers of 30 hidden units, and again we use $\mathrm{tanh}$ activations throughout. Simulations were run in batches of $50$ for the SNL scheme after an initial Fisher pre-training step to initialize the networks.
\subsection{Results}
\begin{figure*}
  \centering
  \subfigure[]{\includegraphics[scale=0.48]{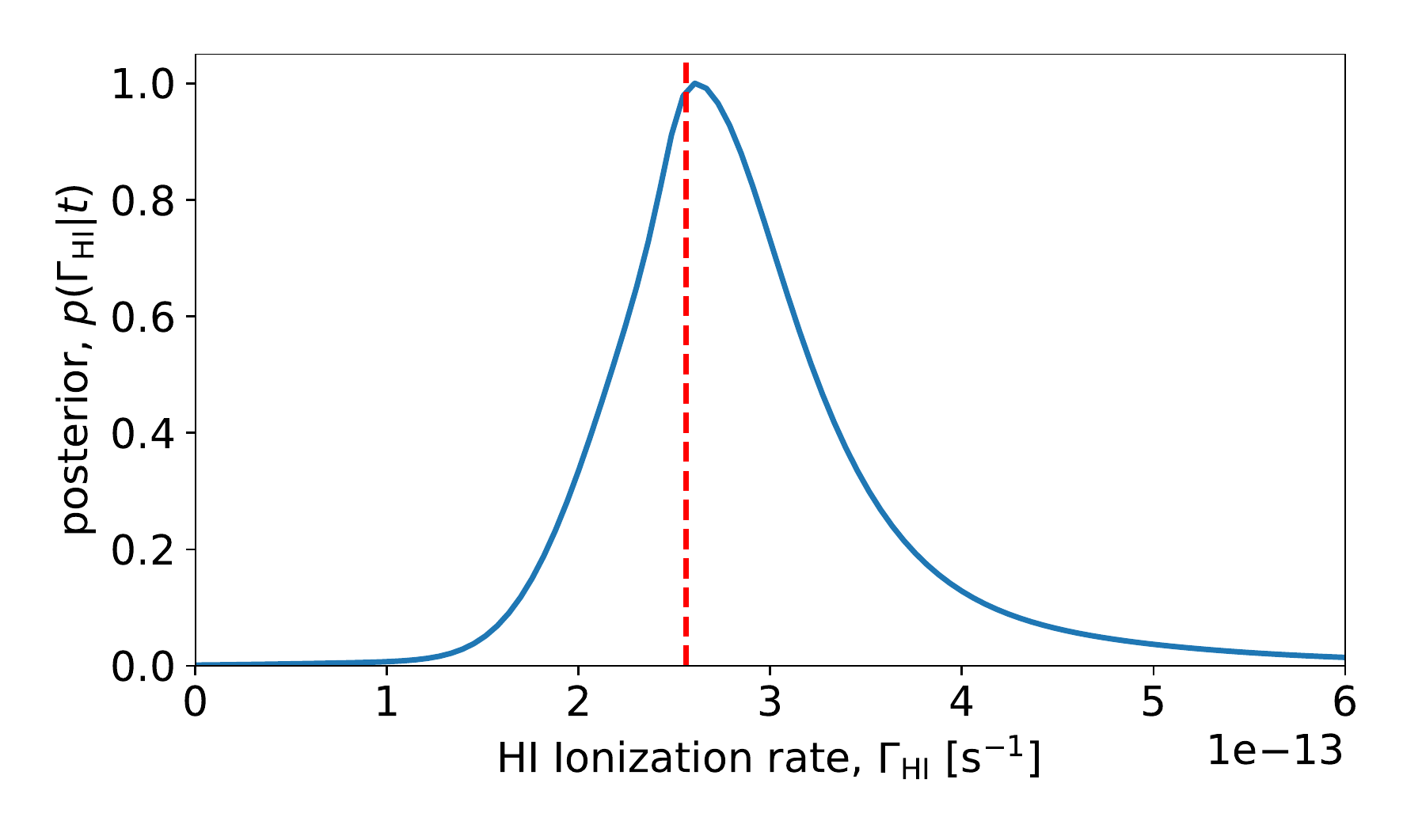}}\quad
  \subfigure[]{\includegraphics[scale=0.48]{gamma_convergence.pdf}}
\caption{Left: Recovered posterior for the ($z=6$) HI ionization rate from the high-$z$ Ly$\alpha$ forest, from DELFI after 300 simulations (mock data shown in Figure \ref{fig:flux_train}). Right: Convergence of the DELFI NDE ensemble for the ionizing background inference task. The DELFI ensemble of NDEs converges extremely quickly in this low-dimensional case, after only $\mathcal{O}(10^2)$ simulations.}
\label{fig:ionizing_background}
\end{figure*}
In Figure \ref{fig:ionizing_background} (right) shows the recovered posterior on the ionization rate $\Gamma_\mathrm{HI}$; the input value (marked in red) is well recovered. We find the DELFI ensemble of neural density estimators converges extremely fast in this case, after only $\mathcal{O}(10^2)$ (Figure \ref{fig:ionizing_background}; left).
\section{Conclusions and discussion}
\label{sec:conclusions}
Density-estimation likelihood-free inference (DELFI) implemented using NDEs to learn the sampling distribution of the data (summaries) as a function of the model parameters, and adaptively acquiring simulations with active learning, provides an efficient framework for likelihood-free inference in cosmology. When combined with massive data compression, high-fidelity posteriors may be achieved from just $\mathcal{O}(10^3)$ forward simulations for typical $\sim 6$ parameter inference tasks. Advances in nuisance-parameter hardened data compression mean that this expected performance may be preserved irrespective of the presence or number of additional nuisance parameters that need to be marginalized over \citep{Alsing2018nuisance}. Even without data compression, DELFI with NDEs and active learning provides a state-of-the-art framework for simulation-based inference (although more simulations will be required for larger, uncompressed data vectors).

We have introduced \textsc{pydelfi} -- a general purpose implementation of DELFI with NDEs and active learning (and data compression) -- available with tutorials and documentation at \url{https://github.com/justinalsing/pydelfi}. \textsc{pydelfi} opens up new possibilities for likelihood-free analyses of complex cosmological data sets, using rich generative models containing physical and observational effects that would otherwise be challenging or impossible to include accurately into a traditional likelihood-based analysis.

For standard inference tasks where the form of the likelihood-function can be assumed known, we note that \textsc{pydelfi} can actually be faster (and more accurate for given resources) than MCMC sampling. By turning the inference problem into a low-dimensional density-estimation task, DELFI effectively builds a fast neural network emulator for the likelihood-function, in a similar spirit to variational inference. We have shown that this can converge quickly, after just $\mathcal{O}(10^3)$ simulations for typical problems, which are typically similar in cost to likelihood evaluations (for simple likelihoods). Meanwhile, MCMC methods would typically require many more likelihood calls to yield well sampled posteriors, for the same number of model parameters. The number of simulations to attain convergence for DELFI in these simple cases may be minimized by using neural density estimators that correspond exactly to the form of the known likelihood, eg., a Gaussian with parameter dependent mean and fixed covariance matrix.

An emerging trend in cosmology is to build emulators for summary statistics for which no robust analytical model exists, such as the non-linear matter power spectrum on small scales \citep{Heitmann2013}, 21cm power spectrum \citep{Schmit2017, Kern2017}, Lyman-$\alpha$ power spectrum \citep{Rogers2018,Bird2018}, weak lensing Minkowski functionals \citep{Marques2018}, and many others. DELFI has a deep connection to emulation methods. Emulators in cosmology have been mostly concerned with learning the expectation value of some summary statistics as a function of the model parameters, which would then typically be plugged into a standard Gaussian likelihood analysis with an estimated covariance matrix. DELFI goes further and builds an emulator for the sampling distribution of the summary statistics, as a function of the model parameters, thereby addressing the expectation emulation and inference tasks in one go and without resorting to restrictive or ad hoc likelihood assumptions in the inference step.

Likelihood-free inference also has a deep connection to Bayesian hierarchical modelling (BHM) approaches to cosmological data analysis. BHMs specify a generative model for the data, which in turn defines a joint likelihood for the hyper-parameters (eg., cosmological and global nuisance parameters) and some latent variables (for example, initial potential fluctuations, true properties and redshifts of individual objects in a survey, etc). These typically high-dimensional likelihoods are then sampled using MCMC (or otherwise), and inference of both the hyper-parameters and latent-variables reported. BHMs and likelihood-free methods are of the same spirit in that they both aim to do inference under as complete a generative model description for the data as possible. However, sampling high-dimensional BHMs for complex forward models is hard and computationally intensive work, and there are often limitations on how rich the implemented models can be in practice. Here likelihood-free methods have a clear advantage over BHMs; simulating forwards is much easier than solving the inverse problem with MCMC sampling or otherwise, and adding extra complexity to the forward model has virtually no impact on the difficulty of the inference task for likelihood-free methods (other than any added cost of running simulations). On the other hand, while likelihood-free methods may rely on data compression to be tractable for high-dimensional data vectors and expensive simulators, sampling methods can target the posterior for the uncompressed data directly and yield inferences of the latent variables as a (potentially useful) by-product.

By relying entirely on forward simulations, the likelihood-free approach marks a shift in the way observational cosmology is done in practice. The scientific effort is reduced to: (1) taking data, (2) building as faithful a forward model and simulation pipeline as possible for those data, and (3) if necessary, devising some data compression scheme to reduce the number of simulations required to achieve accurate posteriors with LFI. Activities that typically make up a large part of traditional cosmological data analysis efforts -- constructing and calibrating intermediate estimators, building and validating approximate likelihoods, computing accurate covariance matrices, etc -- no longer enter into the critical path\footnote{"Critical" in the sense that if they are not done accurately enough, the resulting scientific inferences may be biased.} of scientific reasoning (although they may still be relevant for eg., data compression, but without the same onerous requirements on accuracy as for likelihood-based methods). All critical assumptions underpinning the analysis are then concisely and completely summarized by the forward model specification; this makes for robust science, and clear and simple scientific reporting.
\section*{Acknowledgements}
This work is supported by the Simons Foundation. Justin Alsing was partially supported by the research project grant ``Fundamental Physics from Cosmological Surveys" funded by the Swedish Research Council (VR) under Dnr 2017-04212. Benjamin Wandelt acknowledges support by the Labex Institut Lagrange de Paris (ILP) (reference ANR-10-LABX-63) part of the Idex SUPER, and received financial state aid managed by
the Agence Nationale de la Recherche, as part of the programme Investissements d'avenir
under the reference ANR-11-IDEX-0004-02. Tom Charnock is supported by the ANR BIG4 grant ANR-16-CE23-0002 of the French Agence Nationale de la Recherche and would like to thank NVIDIA for the donation of the Quadro P6000 used in building and testing \textsc{pydelfi}.




\bibliographystyle{mnras}
\bibliography{massive}

\bsp	
\label{lastpage}
\end{document}